\documentclass{article}
\usepackage{amssymb,amsmath}
\usepackage{epsfig}
\usepackage{pdflscape}
\usepackage{subfigure}
\usepackage{color}
\usepackage{lscape}
\usepackage{bbm}
\usepackage{bm}
\usepackage{mathrsfs}
\usepackage{amsfonts}
\usepackage{graphics}
\usepackage{indentfirst}
\usepackage[square,numbers,sort&compress]{natbib}
\usepackage{multicol}

\newtheorem{rem}{Remark}[section]
\newcommand{\br}{\begin{rem}}
\newcommand{\er}{\end{rem}}
\newtheorem{ex}[rem]{Example}
\newcommand{\bex}{\begin{ex}}
\newcommand{\eex}{\end{ex}}
\newtheorem{Def}[rem]{Definition}
\newcommand{\bd}{\begin{Def}}
\newcommand{\ed}{\end{Def}}
\newtheorem{theorem}[rem]{Theorem}
\newcommand{\bt}{\begin{theorem}}
\newcommand{\et}{\end{theorem}}
\newtheorem{Prop}[rem]{Proposition}
\newcommand{\bp}{\begin{Prop}}
\newcommand{\ep}{\end{Prop}}
\newtheorem{lemma}[rem]{Lemma}
\newcommand{\bl}{\begin{lemma}}
\newcommand{\el}{\end{lemma}}
\newcommand{\be}{\begin{equation}}
\newcommand{\ee}{\end{equation}}
\newcommand{\bea}{\begin{eqnarray}}
\newcommand{\eea}{\end{eqnarray}}

\newcommand{\nn}{\nonumber}
\newcommand{\adots}{\mathinner{\mkern2mu\raise1pt\hbox{.}\mkern2mu
\raise4pt\hbox{.}\mkern2mu\raise7pt\hbox{.}\mkern1mu}}

\title{Adding Potentials to Superintegrable Systems with Symmetry}

\author{Allan P. Fordy\thanks{School of Mathematics,
University of Leeds, Leeds LS2 9JT, UK. ~~E-mail: a.p.fordy@leeds.ac.uk}
$\,$ and Qing Huang\thanks{School of Mathematics, Center for Nonlinear Studies, Northwest University, Xi’an 710069,
People’s Republic of China ~~E-mail: hqing@nwu.edu.cn}
}

\begin{document}

\maketitle

\begin{abstract}
In previous work, we have considered Hamiltonians associated with 3 dimensional conformally flat spaces, possessing 2, 3 and 4 dimensional isometry algebras.  Previously our Hamiltonians have represented free motion, but here we consider the problem of adding potential functions in the presence of symmetry.

Separable potentials in the 3 dimensional space reduce to 3 or 4 parameter potentials for Darboux-Koenigs Hamiltonians.  Other 3D coordinate systems reveal connections between Darboux-Koenigs and other well known super-integrable Hamiltonians, such as the Kepler problem and isotropic oscillator.
\end{abstract}



\section{Introduction}

In our previous papers \cite{f19-2,f20-2} we used the method introduced in \cite{f19-3} to construct {\em kinetic energies}, related to 3 dimensional conformally flat spaces and having 4 or 5 independent first integrals.  In both these papers, it is evident that an isometry group for the corresponding metric simplified the construction of a closed Poisson algebra of integrals.

When adding potentials to these kinetic energies, we have the choice of whether or not to impose invariance under some or all of the isometries.  Clearly, imposing more symmetries restricts the potential.  In \cite{f20-2}, we considered isometry algebras of dimension 2, 3 and 4, since there is a ``gap phenomenon'' \cite{14-7} which tells us that 5 symmetries imply 6 symmetries, in which case the space has constant curvature, which we wished to avoid.

In \cite{f20-2}, we were particularly interested in using the isometries to reduce from 3 to 2 dimensions, and showed that for each symmetry algebra, there was a {\em universal reduction}, achieved by adapting coordinates to particular isometries.  In each case, the {\em general} kinetic energy takes {\em separable} form, so is certainly Liouville integrable.  Furthermore, the 2 dimensional metric always has a Killing vector.  In the {\em super-integrable} cases, we can reduce any additional integral which {\em commutes} with the special isometry, being used in the reduction.  The super-integrable reductions are therefore, either constant curvature (possibly flat) or of Darboux-Koenigs type.

For each such reduction we can add a general separable potential and investigate the restrictions imposed by requiring additional integrals.  In the Darboux-Koenigs reductions, the potentials we derive just belong to the classification given in \cite{02-6,03-11}.  However, since our 2 dimensional reductions are, in fact, embedded in a 3 dimensional system, we can choose other coordinate systems, which relate these potentials to other known potentials of super-integrable systems, which can be found in the reviews \cite{90-22,13-2,14-2}.

In particular, curved space resonant oscillators (\ref{2d-e1h4:h4-1rz}) arise in relation to $D_2$ (using the notation of \cite{03-11}).  Curved space generalisations of the Kepler problem arise in (\ref{2d-h1h2:h2-1rHG}) and (\ref{2d-h1h2:h1-2HGr}), in relation to $D_4$ and in (\ref{Kepler-HG-r}), in relation to $D_3$.  In the cases of (\ref{2d-h1h2:h2-1rHG}) and (\ref{Kepler-HG-r}), one of the Darboux-Koenigs first integrals is a generalisation of a component of the Runge-Lenz vector.

In Section \ref{conformal} we give our general notation regarding the 10 dimensional conformal algebra, used throughout.  Section \ref{Sec:GenApp} gives a summary of important results from \cite{f20-2} and an outline of the general approach used in this paper.  Following \cite{f20-2}, we then present our results for individual isometry algebras, in sequence.  Sections \ref{2D-super-reduce-e1h4} and \ref{2D-super-reduce-h1h2} are concerned with the 2D algebras $\left<e_1,h_4\right>$ and $\left<h_1,h_2\right>$, while Sections \ref{2D-super-reduce-e1h1f1} to \ref{2D-super-reduce-h2h3h4} are concerned with the 3D algebras $\left<e_1,h_1,f_1\right>$, $\left<e_1,e_2,h_2\right>$ and $\left<h_2,h_3,h_4\right>$.  In Section \ref{conclude}, we make some general remarks and conclusions.

\section{The 3D Euclidean Metric and its Conformal Algebra}\label{conformal}

In \cite{f20-2}, we considered metrics which are conformally related to the standard Euclidean metric in 3 dimensions, with Cartesian coordinates $(q_1,q_2,q_3)$.  The corresponding kinetic energy takes the form
\be\label{3d-gen}
H = \varphi(q_1,q_2,q_3) \left(p_1^2+p_2^2+p_3^2\right).
\ee
A {\em conformal invariant} $X$, linear in momenta, will satisfy $\{X,H\}=\lambda(X) H$, for some function $\lambda(X)$.  The conformal invariants form a Poisson algebra, which we call the {\em conformal algebra}. In flat spaces, of dimension $n$, the infinitesimal generators consist of $n$ {\em translations}, $\frac{1}{2} n (n-1)$ {\em rotations}, 1 {\em scaling} and $n$ {\em inversions}, totalling $\frac{1}{2} (n+1)(n+2)$.  This algebra is isomorphic to $\mathbf{so}(n+1,1)$  (see Volume $1$, p143, of \cite{84-4}).
For special cases of $\varphi(q_1,q_2,q_3)$ there will be a subalgebra for which $\{X,H\}=0$, thus forming {\em true invariants} of $H$.  These correspond to {\em infinitesimal isometries} (Killing vectors) of the metric.  Constant curvature metrics possess $\frac{1}{2} n(n+1) = 6$ Killing vectors (when $n=3$).

The conformal algebra of this 3D metric has dimension $\frac{1}{2} (n+1)(n+2)=10$ (when $n=3$).  A convenient basis is as follows
\begin{subequations}\label{g1234}
\bea
&&  e_1 = p_1,\quad h_1 = -2(q_1p_1+q_2p_2+q_3p_3),\quad f_1=(q_2^2+q_3^2-q_1^2)p_1-2q_1q_2p_2-2q_1q_3p_3,  \label{g1}\\
&&  e_2 = p_2,\quad h_2=2(q_1p_2-q_2p_1),\quad f_2 =-4q_1q_2p_1-2(q_2^2-q_1^2-q_3^2)p_2-4q_2q_3p_3,  \label{g2}\\
&&  e_3 = p_3,\quad h_3 = 2(q_1p_3-q_3p_1),\quad f_3=-4q_1q_3p_1-4q_2q_3p_2-2(q_3^2-q_1^2-q_2^2)p_3,  \label{g3}\\
&&  h_4 = 4q_3p_2-4q_2p_3.   \label{g4}
\eea
\end{subequations}
The Poisson relations of the ten elements in the conformal algebra (\ref{g1234}) are given in Table \ref{Tab:g1234}.
\begin{table}[h]\centering
\caption{The 10-dimensional conformal algebra (\ref{g1234})}\label{Tab:g1234}\vspace{3mm}
{\footnotesize
\renewcommand\arraystretch{1.26}\begin{tabular}{|c||c|c|c||c|c|c||c|c|c||c|}
\hline   &$e_1$    &$h_1$     &$f_1$   &$e_2$   &$h_2$    &$f_2$    &$e_3$    &$h_3$   &$f_3$    &$h_4$\\[.10cm]\hline\hline
$e_1$    &0        &$2e_1$    &$-h_1$  &0       &$-2e_2$    &$-2h_2$   &0     &$-2e_3$ &$-2h_3$  &0\\[1mm]\hline
$h_1$    &   &0        &$2f_1$  &$-2e_2$ &0        &$2f_2$   &$-2e_3$  &0       &$2f_3$   &0\\[1mm]\hline
$f_1$    &   &   &0       &$-h_2$  &$-f_2$    &0        &$-h_3$  &$-f_3$  &0        &0\\[1mm]\hline\hline
$e_2$    &   &   &   &0       &$2e_1$   &$-2h_1$   &0   & 0       &$h_4$  &$4e_3$\\[1mm]\hline
$h_2$    &   &   &  &   &0        &$-4f_1$         & 0  &$-h_4$   &0   &$4h_3$\\[1mm]\hline
$f_2$    &   &  &   &   &         &0               &$h_4$ &0      &0   &$4f_3$\\[1mm]\hline\hline
$e_3$    &   &   &  &   &   &   &0        &$2e_1$  &$-2h_1$            &$-4e_2$\\[1mm]\hline
$h_3$    &   &   &   &    &   & &    &0       &$-4f_1$                 &$-4h_2$\\[1mm]\hline
$f_3$    &    &  &  &  &   &  &  &  &0                                 &$-4f_2$\\[1mm]\hline\hline
$h_4$    & &  & &   &   &    &    &  &    &0\\[1mm]\hline
\end{tabular}
}\end{table}
Note that this is an example of the conformal algebra given in \cite{f18-1} (Table 3), corresponding to the case $a_2=a_3=2,\; a_4=0$.  The subalgebra $\mathfrak{g}_1$, with basis (\ref{g1}) is just a copy of $\mathfrak{sl}(2)$.  We then make the {\em vector space decomposition} of the full algebra $\mathfrak{g}$ into invariant subspaces under the action of $\mathfrak{g}_1$:
$$
\mathfrak{g} = \mathfrak{g}_1 + \mathfrak{g}_2 + \mathfrak{g}_3 + \mathfrak{g}_4.
$$
The basis elements for $\mathfrak{g}_i$ have the same subscript and are given in the rows of (\ref{g1234}).

\begin{table}[h]
\begin{center}
\caption{The involutions of the conformal algebra (\ref{g1234})}\label{Tab:Invg1234}\vspace{3mm}
{\footnotesize\begin{tabular}{|c||c|c|c||c|c|c||c|c|c||c|}
\hline
         &$e_1$    &$h_1$     &$f_1$   &$e_2$   &$h_2$    &$f_2$  &$e_3$   &$h_3$    &$f_3$   &$h_4$   \\[.10cm]\hline\hline
$\iota_{12}$ &$e_2$ &$h_1$ &$\frac{1}{2}f_2$ &$e_1$ &$-h_2$ &$2 f_1$ &$e_3$ &$-\frac{1}{2}h_4$ &$f_3$ &$-2h_3$  \\[1mm]\hline
$\iota_{13}$ &$e_3$ &$h_1$ &$\frac{1}{2}f_3$ &$e_2$    &$\frac{1}{2}h_4$ &$f_2$ &$e_1$     &$-h_3$ &$2f_1$ &$2h_2$ \\[1mm]\hline
$\iota_{23}$ &$e_1$ &$h_1$ &$f_1$ &$e_3$    &$h_3$ &$f_3$ &$e_2$      &$h_2$ &$f_2$ &$-h_4$ \\[1mm]\hline
$\iota_{ef}$      &$-f_1$ &$-h_1$ &$-e_1$  &$-\frac{1}{2}f_2$ &$h_2$ &$-2e_2$   &$-\frac{1}{2}f_3$ &$h_3$ &$-2e_3$  &$h_4$\\[1mm]\hline
\end{tabular}}
\end{center}
\end{table}

The algebra (\ref{g1234}) possesses a number of involutive automorphisms:
\begin{equation*}
\begin{split}
&  \iota_{12}:\ (q_1,q_2,q_3)\mapsto(q_2,q_1,q_3),\quad \iota_{13}:\ (q_1,q_2,q_3)\mapsto(q_3,q_2,q_1),\quad \iota_{23}:\ (q_1,q_2,q_3)\mapsto(q_1,q_3,q_2),\\
&  \iota_{ef}:\ (q_1,q_2,q_3)\mapsto \left(-\frac{q_1}{q_1^2+q_2^2+q_3^2},-\frac{q_2}{q_1^2+q_2^2+q_3^2},-\frac{q_3}{q_1^2+q_2^2+q_3^2}\right),
\end{split}
\end{equation*}
whose action is given in Table \ref{Tab:Invg1234}.

\subsection{Low Dimensional Subalgebras of the Conformal Algebra $\mathfrak{g}$}\label{sec:subalgebras}

In \cite{f20-2}, we considered subalgebras of $\mathfrak{g}$, having dimensions 2, 3 and 4.  We showed that there were {\em initially} 15 inequivalent subalgebras of interest, but then reduced this to a list of 7 subalgebras, given in Table \ref{Tab:SubAlg-Ham}.

\begin{table}[h]
\begin{center}
\caption{Invariant Hamiltonians for subalgebras of $\mathfrak{g}$}\label{Tab:SubAlg-Ham}\vspace{3mm}
\begin{tabular}{|l|l|l|}
\hline
 Dimension &             Representative             & $\varphi$ of Invariant Hamiltonian    \\[.10cm]\hline
2D   &  $\langle e_1,h_{4}\rangle$ & $\varphi=\psi\left(q_2^2+q_3^2\right)$   \\
  &    $\langle h_1,h_2\rangle$  & $\varphi=(q_1^2+q_2^2)\psi\left(\frac{q_3^2}{q_1^2+q_2^2}\right)$   \\ \hline
3D    &  $\langle e_1,h_1,f_1\rangle$  & $\varphi=q_2^2\psi\left(\frac{q_3}{q_2}\right)$    \\
  &  $\langle e_1,e_2,h_2\rangle$ & $\varphi=\psi(q_3)$    \\
  & $\langle h_2,h_3,h_4\rangle$   & $\varphi=\psi(q_1^2+q_2^2+q_3^2)$   \\ \hline
 4D   & $\langle e_1,h_1,f_1\rangle\oplus\langle h_4\rangle$ &  $\varphi=q_2^2+q_3^2$   \\
 &  $\langle h_2,h_3,h_4\rangle \oplus \langle h_1\rangle$ & $\varphi=q_1^2+q_2^2+q_3^2$   \\  \hline
\end{tabular}
\end{center}
\end{table}

Specifically, we enumerated all such subalgebras, with bases of the form $\left< K_1,\dots ,K_m\right>,\, m=2,3,4$, where $K_i$ are chosen from the list (\ref{g1234}).  A classification of subalgebras of $\mathbf{so}(n+1,1)$ is given in \cite{76-9}, but, since the emphasis is on Lie algebras, this is up to conjugations with respect to the matrix group $O(4,1)$, whereas, in the context of Hamiltonian dynamics, we have the larger group of canonical transformations, so some of their subalgebras can be related.  This classification is used in \cite{15-3}, within the Hamiltonian context, to study a class of ``Superintegrable systems with position dependent mass'', some of which can therefore be related by canonical transformation.

\section{Description of the General Approach}\label{Sec:GenApp}

In \cite{f20-2} we considered the general Hamiltonian $H_0$ (kinetic energy) of the form:
\be\label{3d-genH0}
H_0 = \varphi(q_1,q_2,q_3)\, \left(p_1^2+p_2^2+p_3^2\right),
\ee
where $\varphi(q_1,q_2,q_3)$ is chosen to be one of those listed in Table \ref{Tab:SubAlg-Ham}, corresponding to a particular isometry algebra.

\subsection{Universal Coordinates}\label{univ-coords}

By adapting coordinates to particular symmetries, we then constructed canonical {\em point} transformations $(q_1,q_2,q_3)\mapsto (Q_1,Q_2,Q_3)$, after which $H_0$ took one of two forms:
\begin{subequations}
\bea
H_0 &=& \chi(Q_2) \left(P_1^2+P_2^2+\rho(Q_i) P_3^2\right),  \label{H0-12}  \\
H_0 &=& \chi(Q_2) \left(P_2^2+P_3^2+\rho(Q_i) P_1^2\right),  \label{H0-23}
\eea
having respectively, $P_3$ or $P_1$ as a first integral.  On the level sets of these conserved momenta, the corresponding $H_0$ represents a two dimensional kinetic energy in either $1-2$ or $2-3$ space, with $\rho P_3^2$ and $\rho P_1^2$ being considered as ``potential terms''.  The function $\rho$ is shown to depend upon only \underline{one} of $(Q_1,Q_2)$ in the first case or \underline{one} of $(Q_2,Q_3)$ in the second.  The function $\rho$ is, in fact, totally fixed by the procedure.

\br[Universal Reductions]
The forms of $H_0$ given in (\ref{H0-12}) and (\ref{H0-23}) depend \underline{only} upon the specific symmetry algebra possessed by (\ref{3d-genH0}) and can therefore be considered as {\em universal} for the entire class of metric.  In \cite{f20-2}, we referred to these as {\em universal reductions}.
\er

It is straightforward to add {separable potentials}:
\bea
H &=& \chi(Q_2) \left(P_1^2+P_2^2+\rho(Q_i) (P_3^2+V_3(Q_3))+V_1(Q_1)+V_2(Q_2)\right),  \label{H-12}  \\
H &=& \chi(Q_2) \left(P_2^2+P_3^2+\rho(Q_i) (P_1^2+V_1(Q_1))+V_2(Q_2)+V_3(Q_3)\right),  \label{H-23}
\eea
for three arbitrary functions, $V_\ell(Q_\ell)$, of a \underline{single} variable.  Here it is important that $\rho$ depends upon only one variable.

\medskip
{\em In this paper we restrict to the case when $H$ continues to possess the same {\em conserved momentum}, setting respectively $V_3(Q_3)=c_1$ (in (\ref{H-12})) or $V_1(Q_1)=c_1$ (in (\ref{H-23})).}

\medskip
In each case we evidently have three involutive integrals, which are respectively:
\bea
&& H, \quad P_3,\quad G_1 = P_1^2+\delta_{i1}\, \rho(Q_i) (P_3^2+ c_1)+V_1(Q_1),  \label{H-12-F1}  \\
&& H, \quad P_1,\quad G_1 = P_3^2+\delta_{i3}\, \rho(Q_i) (P_1^2+c_1)+V_3(Q_3),  \label{H-23-F1}
\eea
where $\delta_{ij}$ is the usual Kroneker delta.
\end{subequations}

\subsection{Super-Integrability}

In \cite{f20-2} we also identified {\em specific} forms of the functions $\psi$, of Table \ref{Tab:SubAlg-Ham}, which allowed us to construct quadratic first integrals and thus build super-integrable systems, associated with various kinetic energies $H_0$. This process \underline{fixes} the function $\chi(Q_2)$.  In this paper we deform these ``kinetic'' first integrals with ``potential'' terms and require that they Poisson commute with $H$.

\medskip
In the case of (\ref{H-12}), if $\{H_0,F\}=0$, for some quadratic (in momenta) function $F$, then we require
\be \label{H-12-G2}
\{H,G_2\}=0,\quad\mbox{where}\quad  G_2 = F + \sigma(Q_1,Q_2),
\ee
for some function $\sigma(Q_1,Q_2)$, to be determined.
The bracket $\{H,G_2\}=0$ only contains linear terms in $P_i$, whose coefficients give formulae for the first derivatives of $\sigma$.  This is an overdetermined system, whose integrability conditions impose conditions on the unknown functions $V_1(Q_1),\, V_2(Q_2)$, giving \underline{explicit} forms for each super-integrable case.  These calculations are of the standard type, so we omit most of the details.

\medskip
The case of (\ref{H-23}) is exactly analogous and again gives \underline{explicit} forms for the unknown functions $V_2(Q_2)$, $V_3(Q_3)$ and $\sigma(Q_2,Q_3)$.

\medskip
In these {\em universal coordinates} the super-integrable cases are all of Darboux-Koenigs type with potentials from the classification of \cite{02-6,03-11}.  In other coordinate systems, these take the form of generalisations of some well known super-integrable systems, as mentioned in the introduction.

\subsubsection{Rank of the Integrals}

In \cite{f20-2} we constructed the super-integrable cases in the original Cartesian coordinates $(q_1,q_2,q_3)$.

We found that, in the case of the two dimensional algebras, \underline{all} additional integrals commuted with \underline{one} of the basis elements.  As a consequence, these are essentially 2 dimensional systems, so the rank of the integrals is only 4.  These systems are therefore super-integrable, but \underline{not} maximally.

In the case of the three dimensional algebras, we have a much bigger Poisson algebra, which has rank 5, so these systems \underline{are} maximally super-integrable.  However, in the \underline{universal coordinates}, we can only reduce those integrals which \underline{commute} with the conserved momentum (respectively $P_3$ and $P_1$).  The resulting system is again of rank 4 (including the conserved momentum).

Indeed, in the case of reduction (\ref{H-12}), we have exactly $H, P_3, G_1, G_2$ and $G_3=\{G_1,G_2\}$ (which is {\em cubic} in momenta), with one constraint of the form $G_3^2 = {\cal P}(H, P_3, G_1, G_2)$, which is polynomial of degree 6 in momenta.  The case of reduction (\ref{H-23}) is similar, but with $P_1$ in place of $P_3$.  We only {\em explicitly} present such an algebra once, in (\ref{G1G2G3PB}).

In all these cases (resulting from either 2 or 3 dimensional symmetry algebras) the resulting \underline{2 dimensional} system is maximally super-integrable (rank 3).


\section{Systems with Isometry Algebra $\left<e_1,h_4\right>$}\label{2D-super-reduce-e1h4}

Since this algebra is commutative, we can {\em simultaneously} adapt coordinates to {\em both} basis elements: $\left<e_1,h_4\right>=\left<P_1,P_3\right>$.
The general Hamiltonian in this class is:
\be\label{3d-e1h4}
H = H_0 + U(q_1,q_2,q_3) = \psi\left(q_2^2+q_3^2\right) \left(p_1^2+p_2^2+p_3^2\right) + U(q_1,q_2,q_3).
\ee

We give our {\em universal reductions} (\ref{H-12}) and (\ref{H-23}) for this case, and use (\ref{H-12-G2}) to determine the functions $V_i(Q_i)$ and $\sigma$ for each of our super-integrable cases.

\subsection{Reduction with respect to $h_4$}

The generating function
\begin{subequations}
\be\label{TRe1h4:h4}
S = q_1P_1+\sqrt{q_2^2+q_3^2}\, P_2+\frac14\arctan\left(\frac{q_2}{q_3}\right)P_3,
\ee
gives rise to a Hamiltonian of type (\ref{H-12}):
\be  \label{2d-e1h4:h4}
 H = \psi\left(Q_2^2\right) \left(P_1^2+P_2^2+\frac{P_3^2+c_1}{16 Q_2^2}+V_1(Q_1)+V_2(Q_2)\right),
\ee
where we have set $V_3(Q_3)=c_1$ to retain $P_3=h_4$ as a first integral.

This case corresponds to a conformally flat metric in the $1-2$ space, defined by $P_3 = \mbox{const.}$, with $\frac{P_3^2+c_1}{16 Q_2^2}$ (times the conformal factor) corresponding to a potential term.  Since the conformal factor is a function of only $Q_2$, the momentum $P_1$ corresponds to a Killing vector (in 2D).

We have the following 6 conformal elements:
\bea
&&  {\cal T}_{e_1} = P_1,\;\;\; {\cal T}_{h_1} = -2(Q_1P_1+Q_2P_2),\;\;\; {\cal T}_{f_1}= (Q_2^2-Q_1^2) P_1-2 Q_1Q_2 P_2,  \nn  \\[-1mm]
&&    \label{e1h4-conf-h4}    \\[-1mm]
&&  {\cal T}_{e_2} = P_2,\quad {\cal T}_{h_2} = 2 (Q_1P_2-Q_2 P_1),\quad {\cal T}_{f_2}= 2(Q_1^2-Q_2^2) P_2-4 Q_1Q_2 P_1, \nn
\eea
which satisfy the relations of $\mathfrak{g}_1 + \mathfrak{g}_2$ in Table \ref{Tab:g1234}, together with the algebraic constraints
\be\label{e1h4-conf-constraints}
{\cal T}_{e_1}{\cal T}_{f_1}+\frac{1}{4} {\cal T}_{h_1}^2+\frac{1}{2} \left({\cal T}_{e_2}{\cal T}_{f_2}-\frac{1}{2} {\cal T}_{h_2}^2\right)=0,
                          \quad {\cal T}_{e_1}{\cal T}_{f_2}-2 {\cal T}_{e_2} {\cal T}_{f_1} +{\cal T}_{h_1}{\cal T}_{h_2} =0.
\ee
\end{subequations}

The Hamiltonian (\ref{2d-e1h4:h4}) has the involutive integrals $H, P_3$ and $G_1=P_1^2+V_1(Q_1)$.  For the super-integrable cases, the additional integrals can be written in terms of the conformal elements (\ref{e1h4-conf-h4}).

\subsubsection{The Case $\psi(z)=\frac{z}{\alpha z+\beta}$}\label{sec:e1h4eq1-red}

With this $\psi$, the Hamiltonian (\ref{2d-e1h4:h4}) has the conformal factor of Darboux-Koenigs metric $D_2$.  Apart from $H$ and $P_3$, we have the quadratic integral $G_1=P_1^2+V_1(Q_1)$, from the separation of variables.  We can add a modification of one of our previous integrals to fix the two functions $V_1(Q_1), V_2(Q_2)$.  Each of our previous quadratic functions gives a different class of potential.

\medskip
Adapting our previous function $F_1$, we consider $\{H,G_2\}=0$, where
$$
 G_2={\cal T}_{h_1}^2-4\alpha(Q_1^2+Q_2^2) H+\sigma(Q_1,Q_2),
$$
with ${\cal T}_{h_1}$ given in the list (\ref{e1h4-conf-h4}).  As discussed for (\ref{H-12-G2}), we obtain equations for the derivatives $\sigma_{Q_i}$, whose integrability condition implies restrictions on the functions $V_i$, giving
\begin{subequations}
\be\label{e1h4eq1-V1V2}
  V_1(Q_1)=c_2Q_1^2+\frac{c_3}{Q_1^2},\quad V_2(Q_2)=c_2Q_2^2,\quad \sigma(Q_1,Q_2)=4c_2(Q_1^2+Q_2^2)^2,
\ee
giving
\be \label{2d-e1h4:h4-1}
H = \frac{Q_2^2}{\alpha Q_2^2+\beta}\left(P_1^2+P_2^2+\frac{P_3^2+c_1}{16Q_2^2}+c_2(Q_1^2+Q_2^2)+\frac{c_3}{Q_1^2}\right),
\ee
\end{subequations}
which is the $D_2$ kinetic energy with a potential of ``type B" in the classification of \cite{03-11}.

Adding the cubic integral $G_3=\{G_1,G_2\}$, we find the closed Poisson algebra
\begin{subequations}\label{G1G2G3PB}
\bea
\{G_1,G_3\} &=& 8\alpha H(P_3^2+4G_2-16\beta H)-64G_1G_2+8\alpha(c_1-16c_3) H,  \label{G1G3}\\
\{G_2,G_3\} &=& -32 H(\alpha G_1-2\beta c_2)+32G_1^2-4c_2P_3^2-16c_2G_2-4(c_1+16c_3)c_2,  \label{G2G3} \\
G_3^2 &=& 256\alpha\beta H^2 G_1-16\alpha H(P_3^2G_1+4G_1G_2)+64G_1^2G_2 -256(\alpha^2c_3+\beta^2c_2) H^2  \nn \\
   & &\qquad\  +16 H\left(2\beta c_2P_3^2+8\beta c_2G_2+\alpha(16c_3-c_1)G_1\right)-c_2P_3^2(P_3^2+8G_2)-16c_2G_2^2  \nn \\
    &&  \qquad\ -c_2(c_1-16c_3)(2P_3^2-32\beta H+c_1-16c_3)-8(c_1+16c_3)c_2G_2.   \label{G3square}
\eea
\end{subequations}

\medskip
Adapting our previous function $F_2$, we consider $\{H,G_2\}=0$, where
$$
 G_2=P_1{\cal T}_{h_1}+2\alpha Q_1 H+\sigma(Q_1,Q_2),
$$
with ${\cal T}_{h_1}$ given in the list (\ref{e1h4-conf-h4}).  Again this leads to restrictions on the functions $V_i$, giving
\begin{subequations}
\be\label{e1h4eq12-V1V2}
  V_1(Q_1)=c_2Q_1^2+c_3 Q_1,\quad V_2(Q_2)=\frac{1}{4}\, c_2Q_2^2,\quad \sigma(Q_1,Q_2)= -c_2Q_1(2Q_1^2+Q_2^2)-\frac{c_3}2(4Q_1^2+Q_2^2),
\ee
so
\be \label{2d-e1h4:h4-12}
H = \frac{Q_2^2}{\alpha Q_2^2+\beta}\left(P_1^2+P_2^2+\frac{P_3^2+c_1}{16Q_2^2}+\frac{1}{4}\, c_2\left(4 Q_1^2+Q_2^2\right)+c_3Q_1\right),
\ee
which is the $D_2$ kinetic energy with a potential of ``type A" in the classification of \cite{03-11}.
\end{subequations}
Again, adding the cubic integral $G_3=\{G_1,G_2\}$ leads to a closed Poisson algebra.

\subsubsection*{Other Coordinate Systems}
Using the canonical transformation (\ref{TRe1h4:h4}), we can return to the original Cartesian coordinates. The Hamiltonians (\ref{2d-e1h4:h4-1}) and (\ref{2d-e1h4:h4-12}) then take the forms
\begin{subequations}
\bea
  H &=& \frac{q_2^2+q_3^2}{\alpha(q_2^2+q_3^2)+\beta}\left(p_1^2+p_2^2+p_3^2 +\frac{c_1}{16(q_2^2+q_3^2)}+c_2(q_1^2+q_2^2+q_3^2)+\frac{c_3}{q_1^2}\right),  \label{2d-e1h4:h4-1xy}  \\
  H &=& \frac{q_2^2+q_3^2}{\alpha(q_2^2+q_3^2)+\beta}\left(p_1^2+p_2^2+p_3^2 +\frac{c_1}{16(q_2^2+q_3^2)}+\frac{c_2}4(4q_1^2+q_2^2+q_3^2)+c_3q_1\right), \label{2d-e1h4:h4-12xy}
\eea
\end{subequations}
respectively.  These are extensions of harmonic oscillators, with $h_4$ symmetry, reducing to the flat case when $\beta=0$, being cases 5 and 6 of Table II in \cite{90-22}.

In {\em polar coordinates}, with $(q_1,q_2,q_3) = (r \cos\theta, r \sin\theta \cos\varphi ,r \sin\theta \sin\varphi)$, the case (\ref{2d-e1h4:h4-1xy}) is separable:
\begin{subequations}\label{2d-e1h4:h4-1rz}
\be\label{2d-e1h4:h4-1r}
H = \frac{r^2\sin^2\theta}{\alpha r^2 \sin^2\theta + \beta} \left( p_r^2+\frac{p_\theta^2}{r^2}+\frac{p_\varphi^2}{r^2\sin^2\theta}+\frac{c_1}{16 r^2 \sin^2\theta} + c_2 r^2 + \frac{c_3}{r^2 \cos^2\theta}\right).
\ee

\br[Jacobi's Theorem]
In the initial coordinate system, $(Q_i, P_i)$, $H$ and $G_1$ were simultaneously diagonalised quadratic forms  (in the momenta), since $G_1$ is related to separation in those coordinates. The existence of such quadratic integrals is guaranteed by Jacobi's Theorem. On the other hand, $G_2$ was not related to the separation of variables and is not diagonalised.  In polar coordinates, $G_1$ is no longer diagonal, but $G_2$ is:
$$
G_2 = \frac{4 \beta r^2}{\alpha r^2 \sin^2\theta + \beta} \left(p_r^2-\frac{\alpha \sin^2\theta}{\beta}\, p_\theta^2 - \frac{\alpha p_\varphi^2}{\beta} -\frac{\alpha}{16 \beta} \, (c_1+16 c_3 \tan^2\theta)+c_2\, r^2\right).
$$
Indeed, this can be derived independently, through the separation of variable process.
\er

In {\em cylindrical polar coordinates}, with $(q_1,q_2,q_3) = (z, r \cos\theta,r \sin\theta)$, the case (\ref{2d-e1h4:h4-12xy}) is separable:
\be\label{2d-e1h4:h4-12z}
H = \frac{r^2}{\alpha r^2 + \beta} \left(p_r^2+\frac{p_\theta^2}{r^2}+p_z^2+\frac{c_1}{16 r^2} + \frac{1}{4}\, c_2 r^2 + c_2 z^2+c_3 z\right).
\ee
\end{subequations}
Since $Q_1=z$, the integral $G_1$ remains diagonal.

\subsection{Reduction with respect to $e_1$}

The generating function
\begin{subequations}
\be\label{TRe1h4:e1}
S = q_1P_1+\frac18\log\left(q_2^2+q_3^2\right)\, P_2+\frac14\arctan\left(\frac{q_2}{q_3}\right)P_3,
\ee
gives rise to a Hamiltonian of type (\ref{H-23}):
\be  \label{2d-e1h4:e1}
 H = \frac{1}{16}\, e^{-8 Q_2}\, \psi\left(e^{8 Q_2}\right) \left(P_2^2+P_3^2+ 16 e^{8 Q_2} (P_1^2+c_1)+V_2(Q_2)+V_3(Q_3)\right),
\ee
where we have set $V_1(Q_1)=c_1$ to retain $P_1=e_1$ as a first integral.

This case corresponds to a conformally flat metric in the $2-3$ space, defined by $P_1 = \mbox{const.}$, with $16 e^{8 Q_2} (P_1^2+c_1)$ (times the conformal factor) corresponding to a potential.  Since the conformal factor is a function of only $Q_2$, the momentum $P_3$ corresponds to a Killing vector (in 2D).

We have the following 6 conformal elements:
\bea
&&  {\cal T}_{e_1} = \frac{1}{4} e^{4Q_2} (P_2 \sin 4Q_3-P_3 \cos 4Q_3),\;\;\; {\cal T}_{h_1} = \frac{1}{2} P_2,\;\;\; {\cal T}_{f_1}= -\frac{1}{4} e^{-4Q_2} (P_2 \sin 4Q_3+P_3 \cos 4Q_3), \nn   \\[-1mm]
&&    \label{e1h4-conf-e1}    \\[-1mm]
&&  {\cal T}_{e_2} = \frac{1}{4} e^{4Q_2} (P_2 \cos 4Q_3+P_3 \sin 4Q_3),\;\;\; {\cal T}_{h_2} = -\frac{1}{2} P_3,\;\;\; {\cal T}_{f_2}= \frac{1}{2} e^{-4Q_2} (P_3 \sin 4Q_3-P_2 \cos 4Q_3),  \nn
\eea
which satisfy the relations of $\mathfrak{g}_1 + \mathfrak{g}_2$ in Table \ref{Tab:g1234}, together with the algebraic constraints (\ref{e1h4-conf-constraints}).
\end{subequations}

The Hamiltonian (\ref{2d-e1h4:e1}) has the involutive integrals $H, P_1$ and $G_1=P_3^2+V_3(Q_3)$.  For the super-integrable cases, the additional integrals can be written in terms of the conformal elements (\ref{e1h4-conf-e1}).

\subsubsection{The Case $\psi(z)=\frac{1}{\alpha z+\beta}$}\label{sec:e1h4eq2-red}

With this $\psi$, the Hamiltonian (\ref{2d-e1h4:e1}) has the conformal factor of Darboux-Koenigs metric $D_3$.  Apart from $H$ and $P_1$, we have the quadratic integral $G_1=P_3^2+V_3(Q_3)$, from the separation of variables.  We can add a modification of one of our previous integrals to fix the functions $V_2(Q_2), V_3(Q_3)$ and $\sigma(Q_2,Q_3)$ (of (\ref{H-12-G2})).

\medskip
Adapting our previous function $F_1$, we obtain
\begin{subequations}
\bea
H &=& \frac{e^{-8Q_2}}{16(\alpha e^{8Q_2}+\beta)}\left(P_2^2+P_3^2+16 e^{8Q_2} (P_1^2+c_1) +\frac{c_2+c_3\sin{(8Q_3)}}{\cos^2{(8Q_3)}}\right),  \label{2d-e1h4:e1-1}  \\
 G_2 &=& -\,\frac{1}{2}{\cal T}_{f_1} {\cal T}_{f_2} +\frac{\alpha}{2} e^{8Q_2}\sin{(8Q_3)} H + \frac{e^{-8Q_2}(c_2\sin{(8Q_3)}+c_3)}{32\cos^2{(8Q_3)}},  \label{2d-e1h4:e1-1G2}
\eea
which is the $D_3$ kinetic energy with a potential of ``type B" in the classification of \cite{03-11}.
\end{subequations}

\medskip
Adapting our previous function $F_2 = \frac{1}{4}{\cal T}^2_{f_2}-{\cal T}^2_{f_1} -\alpha e^{8Q_2}\cos{(8Q_3)}H$ leads to an equivalent Hamiltonian.

\subsubsection*{Cartesian coordinates}

Using (\ref{TRe1h4:e1}) to return to the original Cartesian coordinates, (\ref{2d-e1h4:e1-1}) takes the form
\be\label{2d-e1h4:e1-1xy}
  H=\frac1{\alpha(q_2^2+q_3^2)+\beta}\left(p_1^2+p_2^2+p_3^2 +c_1+\frac{c_2+c_3}{32(q_2-q_3)^2}+\frac{c_2-c_3}{32(q_2+q_3)^2}\right),
\ee
which can be further simplified by a rotation of $\frac{\pi}{4}$ in the $q_2 - q_3$ space.

\subsubsection{The Case $\psi(z)=\frac{\sqrt{z}}{\alpha\sqrt{z}+\beta}$}\label{sec:e1h4eq3-red}

With this $\psi$, the Hamiltonian (\ref{2d-e1h4:e1}) has the conformal factor of Darboux-Koenigs metric $D_3$.  Apart from $H$ and $P_1$, we have the quadratic integral $G_1=P_3^2+V_3(Q_3)$, from the separation of variables.  We can add a modification of one of our previous integrals to fix the functions $V_2(Q_2), V_3(Q_3)$ and $\sigma(Q_2,Q_3)$ (of (\ref{H-12-G2})).

\medskip
Adapting our previous function $F_1$, leads to
\begin{subequations}
\bea
 H &=& \frac{\text{e}^{-4Q_2}}{16(\alpha \text{e}^{4Q_2}+\beta)}\left(P_2^2+P_3^2+16\text{e}^{8Q_2}(P_1^2+c_1) +\frac{c_2\sin{(4Q_3)}+c_3}{\cos^2{(4Q_3)}}\right),  \label{2d-e1h4:e1-2}  \\
G_2 &=& -\,\frac{1}{2}\,P_3{\cal T}_{f_2}+2\beta\sin{(4Q_3)} H+\frac{c_2}{8}e^{-4Q_2}-\frac{e^{-4Q_2}\left(c_2+c_3\sin{(4Q_3)}\right)}{4\cos^2{(4Q_3)}},  \label{2d-e1h4:e1-2G2}
\eea
which is the $D_3$ kinetic energy with a potential of ``type B" in the classification of \cite{03-11}.
\end{subequations}

\medskip
Adapting our previous function $F_2 = P_3{\cal T}_{f_1}+2\beta\cos{(4Q_3)} H$, a similar calculation leads to an equivalent Hamiltonian.

\subsubsection*{Other Coordinate Systems}

Using (\ref{TRe1h4:e1}) to return to the original Cartesian coordinates, (\ref{2d-e1h4:e1-2}) takes the form
\be\label{2d-e1h4:e1-2xy}
  H=\frac{\sqrt{q_2^2+q_3^2}}{\alpha\sqrt{q_2^2+q_3^2}+\beta}\left(p_1^2+p_2^2+p_3^2 +c_1+\frac{c_2q_2}{16q_3^2\sqrt{q_2^2+q_3^2}}+\frac{c_3}{16q_3^2}\right).
\ee
When $\beta=0$, this can be found in Table II (case 4) of \cite{90-22}.

In cylindrical coordinates $(q_1,q_2,q_3)=\left(z,r \cos \theta,r \sin \theta\right)$, the Hamiltonian (\ref{2d-e1h4:e1-2xy}) takes the form
\be\label{2d-e1h4:e1-2rz}
  H=\frac{r}{\alpha r+\beta}\left(p_r^2+\frac{p_\theta^2}{r^2}+p_z^2 +c_1+\frac{c_2 \cos\theta}{16 r^2 \sin^2\theta}+\frac{c_3}{16 r^2 \sin^2\theta}\right).
\ee
In these coordinates, $(e_1,h_4) = (p_z,-4 p_\theta)$.


\section{Systems with Isometry Algebra $\left<h_1,h_2\right>$}\label{2D-super-reduce-h1h2}

Since this algebra is commutative, we can {\em simultaneously} adapt coordinates to {\em both} basis elements: $\left<h_1,h_2\right>=\left<P_1,P_3\right>$.
The general Hamiltonian in this class is:
\be\label{3d-h1h2}
H = H_0 + U(q_1,q_2,q_3) = (q_1^2+q_2^2)\psi\left(\frac{q_3^2}{q_1^2+q_2^2}\right)\, \left(p_1^2+p_2^2+p_3^2\right) + U(q_1,q_2,q_3).
\ee
We give our {\em universal reductions} (\ref{H-12}) and (\ref{H-23}) for this case, and use (\ref{H-12-G2}) to determine the functions $V_i(Q_i)$ and $\sigma$ for each of our super-integrable cases.

\subsection{Reduction with respect to $h_2$}

The generating function
\begin{subequations}
\be\label{TRh1h2:h2}
S =-\frac{1}{4}\,\log{\left(q_1^2+q_2^2+q_3^2\right)}\,P_1+\frac12\arctan\left(\frac{q_3}{\sqrt{q_1^2+q_2^2}}\right)\,P_2-\frac12\arctan\left(\frac{q_1}{q_2}\right)\,P_3,
\ee
gives a Hamiltonian of type (\ref{H-12}):
\be   \label{2d-h1h2:h2}
 H = \frac{1}{4}\, \cos^2(2 Q_2)\,\psi\left(\tan^2 2 Q_2\right) \left(P_1^2+P_2^2+\sec^2(2 Q_2) (P_3^2+c_1)+V_1(Q_1)+V_2(Q_2)\right),
\ee
where we have set $V_3(Q_3)=c_1$ to retain $P_3=h_2$ as a first integral.

This case corresponds to a conformally flat metric in the $1-2$ space, defined by $P_3 = \mbox{const.}$, with $\sec^2(2 Q_2) (P_3^2+c_1)$ (times the conformal factor) corresponding to a potential term.  Since the conformal factor is a function of only $Q_2$, the momentum $P_1$ corresponds to a Killing vector (in 2D).

We have the following 6 conformal elements:
\bea
&&  {\cal T}_{e_1} = \frac{1}{2} e^{2Q_1} (P_2 \cos 2 Q_2-P_1 \sin 2Q_2),\quad {\cal T}_{h_1} = P_1,\quad {\cal T}_{f_1}=\frac{1}{2} e^{-2Q_1} (P_2 \cos 2 Q_2+P_1 \sin 2Q_2), \nn  \\[-1mm]
&&       \label{h1h2-confh2}   \\[-1mm]
&&  {\cal T}_{e_2} = \frac{1}{2} e^{2Q_1} (P_2 \sin 2 Q_2+P_1 \cos 2Q_1),\quad {\cal T}_{h_2} = P_2,\quad {\cal T}_{f_2}=  e^{-2Q_1} (P_2 \sin 2 Q_2-P_1 \cos 2Q_2),  \nn
\eea
which satisfy the relations of $\mathfrak{g}_1 + \mathfrak{g}_2$ in Table \ref{Tab:g1234}, together with the algebraic constraints (\ref{e1h4-conf-constraints}).
\end{subequations}

The Hamiltonian (\ref{2d-h1h2:h2}) has the involutive integrals $H, P_3$ and $G_1 = P_1^2+V_1(Q_1)$.  For the super-integrable cases, the additional integrals can be written in terms of the conformal elements (\ref{h1h2-confh2}).

\subsubsection{The Case $\psi(z)=\frac{\sqrt{1+z}}{\alpha \sqrt{1+z} +\beta \sqrt{z}}$}\label{sec:2d-h1h2:h2-1}

With this $\psi$, the Hamiltonian (\ref{2d-h1h2:h2}) has the conformal factor of Darboux-Koenigs metric $D_4$.  Apart from $H$ and $P_3$, we have the quadratic integral $G_1=P_1^2+V_1(Q_1)$, from the separation of variables.  We can add a modification of one of our previous integrals to fix the functions $V_1(Q_1), V_2(Q_2)$ and $\sigma(Q_1,Q_2)$ (of (\ref{H-12-G2})).

\medskip
Adapting our previous function $F_1$, we consider $\{H,G_2\}=0$, and obtain
\begin{subequations}
\bea
H &=& \frac{\cos^2{(2Q_2)}}{4(\alpha+\beta\sin{(2Q_2)})} \left(P_1^2+P_2^2+\sec^2{(2Q_2)} (P_3^2+c_1) +c_2 e^{-4Q_1}+c_3 e^{-2Q_1}\right),  \label{2d-h1h2:h2-1}  \\
G_2 &=& P_1{\cal T}_{e_1}-\beta e^{2Q_1} H  -\frac{1}{4}\, \left(2 c_2 e^{-2Q_1}+c_3\right)\sin{(2Q_2)},   \label{2d-h1h2:h2-1G2}
\eea
which is the $D_4$ kinetic energy with a potential of ``type A" in the classification of \cite{03-11}.
\end{subequations}

\medskip
Adapting our previous function $F_2 = 2P_1{\cal T}_{f_1} +2\beta\text{e}^{-2Q_1} H$, we find an equivalent Hamiltonian, given by $Q_1\mapsto -Q_1$.

\subsubsection*{Other Coordinate Systems}

Using (\ref{TRh1h2:h2}) to return to the original Cartesian coordinates, (\ref{2d-h1h2:h2-1}) takes the form
\begin{subequations}
\be\label{2d-h1h2:h2-1qp}
  H=\frac{\sqrt{q_1^2+q_2^2+q_3^2}\, (q_1^2+q_2^2)}{\alpha\sqrt{q_1^2+q_2^2+q_3^2}+\beta q_3}\left(p_1^2+p_2^2+p_3^2 +\frac{c_1}{4(q_1^2+q_2^2)}+\frac{c_2}4+\frac{c_3}{4\sqrt{q_1^2+q_2^2+q_3^2}}\right).
\ee
This is a conformally flat extension of case 7 in Table II of \cite{90-22}, constant {\em scalar} curvature when $\beta=0$.  In these coordinates
\bea
G_1 &=& h_1^2 +c_2 (q_1^2+q_2^2+q_3^2)^2 + c_3 \sqrt{q_1^2+q_2^2+q_3^2},  \label{2d-h1h2:h2-1G1qp}  \\
G_2 &=& e_3 h_1 - \frac{\beta H}{\sqrt{q_1^2+q_2^2+q_3^2}} -\frac{q_3}{4} \left(2 c_2+\frac{c_3}{\sqrt{q_1^2+q_2^2+q_3^2}}\right).  \label{2d-h1h2:h2-1G2qp}
\eea
\end{subequations}

In {\em polar coordinates}, with $(q_1,q_2,q_3) = (r \sin\theta \cos\varphi ,r \sin\theta \sin\varphi, r \cos\theta)$, we have
\begin{subequations}\label{2d-h1h2:h2-1rHG}
\be\label{2d-h1h2:h2-1r}
  H=\frac{r^2 \sin^2\theta}{\alpha +\beta \cos\theta}\left(p_r^2+\frac{p_\theta^2}{r^2}+\frac{p_\varphi^2}{r^2\sin^2\theta} +\frac{c_1}{4 r^2 \sin^2\theta}+\frac{c_2}{4}+\frac{c_3}{4 r}\right),
\ee
with $(h_1,h_2) = (-2 r p_r,2 p_\varphi)$ and
\bea
G_1 &=& 4 r^2 p_r^2 +c2 r^2 + c_3 r,  \label{2d-h1h2:h2-1G1r}  \\
G_2 &=& 2 p_r (p_\theta \sin\theta-r p_r \cos\theta) - \frac{\beta H}{r} -\frac{1}{4} (c_3+2 c_2 r)\cos\theta.  \label{2d-h1h2:h2-1G2r}
\eea
\end{subequations}
It can be seen that this is an extension of the Kepler problem (but neither flat nor fully rotationally invariant).  The first integral $G_2$ is an extension of one component of the Runge-Lenz vector, which is most easily seen in the Cartesian form (\ref{2d-h1h2:h2-1G2qp}).  The Hamiltonian (\ref{2d-h1h2:h2-1r}) is separable in these coordinates.

\subsubsection{The Case $\psi(z)=\frac{z}{\alpha +\beta z}$}\label{sec:2d-h1h2:h2-2}

With this $\psi$, the Hamiltonian (\ref{2d-h1h2:h2}) has the conformal factor of Darboux-Koenigs metric $D_4$.  Apart from $H$ and $P_3$, we have the quadratic integral $G_1=P_1^2+V_1(Q_1)$, from the separation of variables.  We can add a modification of one of our previous integrals to fix the functions $V_1(Q_1), V_2(Q_2)$ and $\sigma(Q_1,Q_2)$ (of (\ref{H-12-G2})).

\medskip
Adapting our previous function $F_1$, we consider $\{H,G_2\}=0$, and find
\begin{subequations}
\bea
H &=& \frac{\sin^2{(4Q_2)}}{8((\alpha-\beta)\cos{(4Q_2)}+\alpha+\beta)}   \left(P_1^2+P_2^2+\sec^2{(2Q_2)} (P_3^2+c_1)+c_2e^{-8Q_1}+c_3e^{-4Q_1}\right),  \label{2d-h1h2:h2-2}  \\
G_2 &=& {\cal T}_{e_1}^2-\frac{\alpha e^{4Q_1}}{\sin^2{(2Q_2)}} H+ \frac{c_2}{4} e^{-4Q_1}\sin^2{(2Q_2)},  \label{2d-h1h2:h2-2G2}
\eea
which is the $D_4$ kinetic energy with a potential of ``type A" in the classification of \cite{03-11}.
\end{subequations}

\medskip
Adapting our previous function $F_2 = 4{\cal T}_{f_1}^2-\frac{4\alpha\text{e}^{-4Q_1}}{\sin^2{(2Q_2)}} H$, leads to an equivalent Hamiltonian, corresponding to $Q_1\mapsto -Q_1$.

\br[Variations on the ``type'']\label{rem:variations}
The potential of (\ref{2d-h1h2:h2-2}) differs from the true ``type A'' in the term $\sec^2{(2Q_2)}$, which replaces the term $\csc^2{(4Q_2)}$ in the classification of \cite{03-11} (taking into account a rescaling).  In fact the classification of \cite{03-11} is very dependent on the choice of integrals being used.  In (\ref{2d-h1h2:h2-1G2}), we use $P_1{\cal T}_{e_1}$ as the leading term in $G_2$ (coinciding with the choice in \cite{03-11}), so obtain \underline{exactly} their potential.  In (\ref{2d-h1h2:h2-2G2}), we have used ${\cal T}_{e_1}^2$, which leads to this small variation.  In fact, we are not \underline{choosing} these integrals at this stage, since they just arise through the reduction process from some integrals, which were \underline{previously} chosen in \cite{f20-2}.
\er

\subsubsection*{Cartesian coordinates}

In the original Cartesian coordinates, (\ref{2d-h1h2:h2-2}) is of the form
\be\label{2d-h1h2:h2-2xy}
  H=\frac{(q_1^2+q_2^2)\, q_3^2}{\alpha(q_1^2+q_2^2)+\beta q_3^2}\left(p_1^2+p_2^2+p_3^2+\frac{c_1}{4(q_1^2+q_2^2)}+\frac{c_2}{4}(q_1^2+q_2^2+q_3^2) +\frac{c_3}{4}\right).
\ee
This is a conformally flat extension of case 5 in Table II of \cite{90-22}, constant {\em scalar} curvature when $\beta=0$.

\subsection{Reduction with respect to $h_1$}

The generating function
\begin{subequations}
\be\label{TRh1h2:h1}
S = -\frac{1}{4}\,\log\left(q_1^2+q_2^2+q_3^2\right)\,P_1+\frac12\, \log\left(\frac{q_3+\sqrt{q_1^2+q_2^2+q_3^2}}{\sqrt{q_1^2+q_2^2}}\right)\,P_2-\frac12\arctan\left(\frac{q_1}{q_2}\right)\,P_3,
\ee
gives a Hamiltonian of type (\ref{H-23}):
\be  \label{2d-h1h2:h1}
 H = \frac{1}{4}\,  \psi\left(\sinh^2 2 Q_2\right) \left(P_2^2+P_3^2+ \mbox{sech}^2\,(2 Q_2)\,  (P_1^2+c_1)+V_2(Q_2)+V_3(Q_3)\right),
\ee
where we have set $V_1(Q_1)=c_1$ to retain $P_1=h_1$ as a first integral.

This case corresponds to a conformally flat metric in the $2-3$ space, defined by $P_1 = \mbox{const.}$, with $\mbox{sech}^2\,(2 Q_2)\,  (P_1^2+c_1)$ (times the conformal factor) corresponding to a potential term.  Since the conformal factor is a function of only $Q_2$, the momentum $P_3$ corresponds to a Killing vector (in 2D).

We have the following 6 conformal elements:
\bea
&&  {\cal T}_{e_1} = \frac{1}{2} e^{2Q_2} (P_2 \sin 2 Q_3-P_3 \cos 2Q_3),\quad {\cal T}_{h_1} = P_2,\quad {\cal T}_{f_1}= -\frac{1}{2} e^{-2Q_2} (P_2 \sin 2 Q_3+P_3 \cos 2Q_3), \nn  \\[-1mm]
&&       \label{h1h2-confh1}   \\[-1mm]
&&  {\cal T}_{e_2} = \frac{1}{2} e^{2Q_2} (P_2 \cos 2 Q_3+P_3 \sin 2Q_3),\quad {\cal T}_{h_2} = -P_3,\quad {\cal T}_{f_2}=  e^{-2Q_2} (P_3 \sin 2Q_3- P_2 \cos 2 Q_3),  \nn
\eea
which satisfy the relations of $\mathfrak{g}_1 + \mathfrak{g}_2$ in Table \ref{Tab:g1234}, together with the algebraic constraints (\ref{e1h4-conf-constraints}).
\end{subequations}

The Hamiltonian (\ref{2d-h1h2:h1}) has the involutive integrals $H, P_1$ and $G_1=P_3^2+V_3(Q_3)$.  For the super-integrable cases, the additional integrals can be written in terms of the conformal elements (\ref{h1h2-confh1}).

\subsubsection{The Case $\psi(z)=\frac{(1+z) z}{\alpha +\beta z}$}\label{sec:2d-h1h2:h1-1}

With this $\psi$, the Hamiltonian (\ref{2d-h1h2:h1}) has the conformal factor of Darboux-Koenigs metric $D_4$.  Apart from $H$ and $P_1$, we have the quadratic integral $G_1=P_3^2+V_3(Q_3)$, from the separation of variables.  We can add a modification of one of our previous integrals to fix the functions $V_2(Q_2), V_3(Q_3)$ and $\sigma(Q_2,Q_3)$ (of (\ref{H-12-G2})).

\medskip
Adapting our previous function $F_2$, we find
\begin{subequations}
\bea
  H &=& \frac{\sinh^2{(4Q_2)}}{8(\beta\cosh{(4Q_2)}+2\alpha-\beta)}\left(P_2^2+P_3^2+\mbox{sech}^2\,(2 Q_2)\, (P_1^2+c_1) \right. \nn\\
  &&   \hspace{6cm}  \left.  +\frac{c_2}{4\sinh^2{(4Q_2)}} +\frac{c_3\cos{(4Q_3)}+c_4}{\sin^2{(4Q_3)}}\right),                       \label{2d-h1h2:h1-1}  \\
 G_2 &=& 4({\cal T}_{e_1}-{\cal T}_{f_1})^2-(2{\cal T}_{e_2}-{\cal T}_{f_2})^2+\frac{16\alpha\cos{(4Q_3)}}{\sinh^2{(2Q_2)}} H
              -\frac{c_2\cos{(4Q_3)}}{4\sinh^2{(2Q_2)}}+\frac{4c_3\sinh^2{(2Q_2)}}{\sin^2{(4Q_3)}},                           \label{2d-h1h2:h1-1G2}
\eea
\end{subequations}
which is the $D_4$ kinetic energy with a potential of ``type B" in the classification of \cite{03-11}. Apart from the switch of trigonometric and hyperbolic functions, this also has a slight variation of the type described in Remark \ref{rem:variations}.

\medskip
Adapting our previous function $F_1 = ({\cal T}_{e_1}-{\cal T}_{f_1})(2{\cal T}_{e_2}-{\cal T}_{f_2})-\frac{4\alpha\sin{(4Q_3)}}{\sinh^2{(2Q_2)}} H$, we obtain an equivalent Hamiltonian.

\subsubsection*{Cartesian coordinates}

In the original Cartesian coordinates, (\ref{2d-h1h2:h1-1}) is of the form
\be\label{2d-h1h2:h1-1xy}
  H=\frac{(q_1^2+q_2^2+q_3^2)\, q_3^2}{\alpha(q_2^2+q_3^2)+\beta q_3^2}\left(p_1^2+p_2^2+p_3^2 +\frac{16c_1-c_2}{64(q_1^2+q_2^2+q_3^2)} +\frac{1}{16}\,\left(\frac{c_4+c_3}{q_1^2}+\frac{c_4-c_3}{q_2^2}+\frac{c_2}{4 q_3^2}\right)\right).
\ee
This is a conformally flat extension of case 1 in Table II of \cite{90-22}, constant {\em scalar} curvature when $\alpha=0$.

\subsubsection{The Case $\psi(z)=\frac{1+z}{\alpha  +\beta \sqrt{z}}$}\label{sec:2d-h1h2:h1-2}

With this $\psi$, the Hamiltonian (\ref{2d-h1h2:h1}) has the conformal factor of Darboux-Koenigs metric $D_4$.  Apart from $H$ and $P_1$, we have the quadratic integral $G_1=P_3^2+V_3(Q_3)$, from the separation of variables.  We can add a modification of one of our previous integrals to fix the functions $V_2(Q_2), V_3(Q_3)$ and $\sigma(Q_2,Q_3)$ (of (\ref{H-12-G2})).

\medskip
Adapting our previous function $F_1$, we find
\begin{subequations}
\bea
H &=& \frac{\cosh^2{(2Q_2)}}{4(\alpha+\beta\sinh{(2Q_2)})}\left(P_2^2+P_3^2+\mbox{sech}^2\,(2 Q_2)\, (P_1^2+c_1) +\frac{c_2+c_3\cos{(2Q_3)}}{\sin^2{(2Q_3)}}\right),  \label{2d-h1h2:h1-2}  \\
 G_2 &=& P_3\left({\cal T}_{f_1}-{\cal T}_{e_1}\right)-2\beta\cos{(2Q_3)} H + \left(\frac{c_2\cos{(2Q_3)}+c_3}{\sin^2{(2Q_3)}}-\frac{c_3}2\right)\sinh{(2Q_2)},\label{2d-h1h2:h1-2G2}
\eea
which is the $D_4$ kinetic energy with a potential of ``type B" in the classification of \cite{03-11}.
\end{subequations}
Adding the cubic integral $G_3=\{G_1,G_2\}$, we again find a closed Poisson algebra.

\medskip
Adapting our previous function $F_2 = P_3\left({\cal T}_{f_2}-2{\cal T}_{e_2}\right)+4\beta\sin{(2Q_3)} H$, leads to an equivalent Hamiltonian.

\subsubsection*{Other Coordinate Systems}

Using (\ref{TRh1h2:h1}) to return to the original Cartesian coordinates, (\ref{2d-h1h2:h1-2}) takes the form
\begin{subequations}
\be\label{2d-h1h2:h1-2xy}
  H=\frac{\sqrt{q_1^2+q_2^2}\, (q_1^2+q_2^2+q_3^2)}{\alpha\sqrt{q_1^2+q_2^2}+\beta q_3}\left(p_1^2+p_2^2+p_3^2 +\frac{c_1}{4(q_1^2+q_2^2+q_3^2)} +\frac{c_2}{4q_1^2}+\frac{c_3q_2}{4q_1^2\sqrt{q_1^2+q_2^2}}\right),
\ee
and
\be
G_1 = h_2^2+ c_2 \left(\frac{q_1^2+q_2^2}{q_1^2}\right) +\frac{c_3 q_2 \sqrt{q_1^2+q_2^2}}{q_1^2}, \quad
G_2 = h_2 h_3 -\frac{2 \beta q_2}{\sqrt{q_1^2+q_2^2}}\, H + \frac{c_2q_2}{q_1^2} +\frac{c_3 (q_1^2+2 q_2^2)}{2 q_1^2 \sqrt{q_1^2+q_2^2}}.  \label{2d-h1h2:h1-2G12xy}
\ee
\end{subequations}

In {\em polar coordinates}, with $(q_1,q_2,q_3) = (r \sin\theta \cos\varphi ,r \sin\theta \sin\varphi, r \cos\theta)$, we have
\begin{subequations}\label{2d-h1h2:h1-2HGr}
\be\label{2d-h1h2:h1-2r}
  H=\frac{r^2 \sin\theta}{\alpha\sin\theta +\beta \cos\theta}\left(p_r^2+\frac{p_\theta^2}{r^2}+\frac{p_\varphi^2}{r^2\sin^2\theta} +\frac{c_1}{4 r^2}+\frac{\sec^2\varphi}{4r^2 \sin^2\theta}\, (c_2+c_3 \sin\varphi)\right),
\ee
with $(h_1,h_2) = (-2 r p_r,2 p_\varphi)$ and
\bea
G_1 &=& 4 p_\varphi^2 +\sec^2\varphi (c_2+c_3 \sin\varphi),  \label{2d-h1h2:h1-2G1r}  \\
G_2 &=& 4 p_\varphi (p_\varphi \cot\theta\sin\varphi - p_\theta \cos\varphi) - 2\beta\sin\varphi H +2 c_2 \sec^2\varphi \sin \varphi+c_3 (1+2 \tan^2 \varphi).  \label{2d-h1h2:h1-2G2r}
\eea
\end{subequations}
It can be seen that this is an extension of the Kepler problem (but neither flat nor fully rotationally invariant).  When $\beta=c_2=c_3=0$, the Hamiltonian \underline{is} fully rotationally invariant, with $G_1$ and $G_2$ reducing to products of angular momenta.  The Hamiltonian (\ref{2d-h1h2:h1-2r}) is separable in these coordinates.


\section{Systems with Isometry Algebra $\langle e_1,h_1,f_1\rangle$}\label{2D-super-reduce-e1h1f1}

In this (non-commutative) case we adapt coordinates to either $e_1\rightarrow P_1$ \underline{or} $h_1\rightarrow P_3$.  The case of $f_1$ is related to that of $e_1$ through the involution $\iota_{ef}$.
The general Hamiltonian in this class is:
\be\label{3d-e1h1f1}
H = H_0 + U(q_1,q_2,q_3) = q_2^2 \psi\left(\frac{q_3}{q_2}\right)\, \left(p_1^2+p_2^2+p_3^2\right) + U(q_1,q_2,q_3).
\ee

\subsection{Reduction with respect to $h_1\mapsto P_3$}

The generating function
\begin{subequations}
\be\label{e1h1f1-S3h1}
  S=\frac{1}{2}\, \log\left(\frac{\sqrt{q_1^2+q_2^2+q_3^2}+q_1}{\sqrt{q_1^2+q_2^2+q_3^2}-q_1}\right)\;P_1+\arctan\left(\frac{q_3}{q_2}\right)\;P_2-\frac14\log\left(q_1^2+q_2^2+q_3^2\right)\;P_3,
\ee
gives a Hamiltonian of type (\ref{H-12}):
\be\label{e1h1f1-Ht2h1}
  H=\cos^2{Q_2}\;\psi\left(\tan{Q_2}\right)\left(P_1^2+P_2^2+\frac{1}{4\cosh^2Q_1}(P_3^2+c_1) +V_1(Q_1)+V_2(Q_2)\right),
\ee
where we have set $V_3(Q_3)=c_1$ to retain $P_3=h_1$ as a first integral.

\smallskip
This case corresponds to a conformally flat metric in the $1-2$ space, defined by $P_3 = \mbox{const.}$, with $\frac{1}{4\cosh^2Q_1}(P_3^2+c_1)$ (times the conformal factor) corresponding to a potential term.  Since the conformal factor is a function of only $Q_2$, the momentum $P_1$ corresponds to a Killing vector (in 2D).

We have the following 6 conformal elements:
\bea
&&  {\cal T}_{e_1} = e^{Q_1} (P_2 \sin Q_2+P_1 \cos Q_2),\quad {\cal T}_{h_1} = 2 P_1,\quad {\cal T}_{f_1}=e^{-Q_1} (P_2 \sin Q_2-P_1 \cos Q_2), \nn  \\[-1mm]
&&                                                            \label{e1h1f1-conf-h1}    \\[-1mm]
&&  {\cal T}_{e_2} = e^{Q_1} (P_2 \cos Q_2-P_1 \sin Q_2),\quad {\cal T}_{h_2} = -2 P_2,\quad {\cal T}_{f_2}= 2e^{-Q_1} (P_2 \cos Q_2+P_1 \sin Q_2),  \nn
\eea
which satisfy the relations of $\mathfrak{g}_1 + \mathfrak{g}_2$ in Table \ref{Tab:g1234}, together with the algebraic constraints (\ref{e1h4-conf-constraints}).
\end{subequations}

The Hamiltonian (\ref{e1h1f1-Ht2h1}) has the involutive integrals $H, P_3$ and $G_1=P_1^2+\frac{1}{4\cosh^2Q_1}(P_3^2+c_1)+V_1(Q_1)$.  For the super-integrable cases, the additional integrals can be written in terms of the conformal elements (\ref{e1h1f1-conf-h1}).

\subsubsection{The Case $\psi(z)=\frac{z^2}{\alpha+\beta z^2}$}\label{2D-super-reduce-e1h1f1-h1-1}

With this $\psi$, the Hamiltonian (\ref{e1h1f1-Ht2h1}) takes the form given below.  Apart from $H$ and $P_3$, we have the quadratic integrals $G_1, G_2$:
\bea
H &=& \frac{\sin^2(2 Q_2)}{2(\alpha+\beta)+2(\alpha-\beta) \cos(2Q_2)} \left(P_1^2+P_2^2+\frac{1}{4}\, \mbox{sech}^2\, Q_1 \left(P_3^2+c_1\right)+\frac{c_2}{\sin^2{(2Q_2)}} +\frac{c_3}{\sinh^2{Q_1}}\right), \nn   \\
G_1 &=& P_1^2+\frac{1}{4}\, \mbox{sech}^2\, Q_1  \left(P_3^2+c_1\right)+\frac{c_3}{\sinh^2{Q_1}} , \nn  \\
G_2 &=&  \frac{1}{2} ({\cal T}_{e_1}+{\cal T}_{f_1})^2+\beta (\cos 2 Q_2-\cosh 2 Q_1) \sec^2 Q_2\, H -\frac{1}{2}\, (P_3^2+c_1)\, \cos^2 Q_2\, \mbox{sech}^2\, Q_1 +\frac{c_2\cosh^2{Q_1}}{2\cos^2{Q_2}}.  \nn
\eea
This Hamiltonian is the $D_4$ kinetic energy with a potential of ``type B" in the classification of \cite{03-11}.

\subsubsection*{In Cartesian Coordinates}

In the original Cartesian coordinates, the Hamiltonian is of the form
\be\label{2D-super-reduce-e1h1f1-h1-1qp}
H = \frac{q_2^2q_3^2}{\alpha q_2^2+\beta q_3^2}\left(p_1^2+p_2^2+p_3^2 +\frac{c_1}{4(q_1^2+q_2^2+q_3^2)}+\frac{c_2}{4}\left(\frac{1}{q_2^2}+\frac{1}{q_3^2}\right)+\frac{c_3}{q_1^2}\right).
\ee
This is a conformally flat extension of case 1 in Table II of \cite{90-22}, constant curvature when either $\alpha=0$ or $\beta=0$.

\subsubsection{The Case $\psi(z)=\frac{\sqrt{1+z^2}}{\alpha \sqrt{1+z^2}  +\beta z}$}\label{2D-super-reduce-e1h1f1-h1-2}

With this $\psi$, the Hamiltonian (\ref{e1h1f1-Ht2h1}) takes the form of (\ref{2D-super-reduce-e1h1f1-h1-2H}).  Apart from $H$ and $P_3$, we have the quadratic integrals $G_1, G_2$, given below.
\begin{subequations}
\bea
H &=& \frac{\cos^2 Q_2}{\alpha+\beta \sin Q_2} \, \left(P_1^2+P_2^2+\frac{1}{4}\, \mbox{sech}^2\, Q_1 \left(P_3^2+c_1\right) +\frac{c_2 \sinh{Q_1}}{2 \cosh^2{Q_1}}+c_3 \sec^2{Q_2}\right), \label{2D-super-reduce-e1h1f1-h1-2H}\\
G_1 &=& P_1^2+\frac{1}{4}\, \mbox{sech}^2\, Q_1  \left(P_3^2+c_1\right)+\frac{c_2 \sinh{Q_1}}{2 \cosh^2{Q_1}} ,  \label{2D-super-reduce-e1h1f1-h1-2G1} \\
G_2 &=& 4 P_2 \left({\cal T}_{e_1}-{\cal T}_{f_1}\right) + 2  \sec^2 Q_2\,\sinh Q_1 \, \left(\beta \cos{(2 Q_2)}-3 \beta -4 \alpha \sin Q_2\right) H \nn\\
                              &&  \hspace{6cm} + \sin{Q_2} (2c_2+8 c_3\,\sec^2 Q_2\,\sinh Q_1).  \label{2D-super-reduce-e1h1f1-h1-2G2}
\eea
\end{subequations}
This Hamiltonian is the $D_4$ kinetic energy with a potential of ``type B" in the classification of \cite{03-11}.

\subsubsection*{Other Coordinate Systems}

Using (\ref{e1h1f1-S3h1}) to return to the original Cartesian coordinates, the Hamiltonian (\ref{2D-super-reduce-e1h1f1-h1-2H}) takes the form
\begin{subequations}
\be\label{2D-super-reduce-e1h1f1-h1-2qp}
H = \frac{q_2^2\sqrt{q_2^2+q_3^2}}{\alpha\sqrt{q_2^2+q_3^2}+\beta q_3}\left(p_1^2+p_2^2+p_3^2 +\frac{c_1}{4(q_1^2+q_2^2+q_3^2)}+\frac{c_2q_1}{2\sqrt{q_2^2+q_3^2}(q_1^2+q_2^2+q_3^2)}+\frac{c_3}{q_2^2}\right).
\ee
When $\beta=0$, this reduces to a constant curvature case.

This is separable in {\em polar coordinates} $(q_1,q_2,q_3) = (r \cos\theta,r \sin\theta \cos\varphi ,r \sin\theta \sin\varphi)$:
\be\label{2D-super-reduce-e1h1f1-h1-2r}
H = \frac{r^2 \sin^2\theta \cos^2\varphi}{\alpha+\beta \sin\varphi}\left(p_r^2+\frac{p_\theta^2}{r^2}+\frac{p_\varphi^2}{r^2\sin^2\theta}
                                                   +\frac{c_1}{4 r^2}+\frac{c_2 \cot\theta}{2 r^2} +\frac{c_3 \sec^2\varphi}{r^2 \sin^2\theta}\right).
\ee
\end{subequations}

\subsection{Reduction using the Isometry $e_1\mapsto P_1$}

The generating function
\begin{subequations}
\be\label{e1h1f1-S3e1}
  S=q_1\;P_1+\arctan \left(\frac{q_3}{q_2}\right)\;P_2+\frac12\log \left(q_2^2+q_3^2\right)\;P_3,
\ee
gives a Hamiltonian of type (\ref{H-23}):
\be\label{e1h1f1-Ht2e1}
  H =\cos^2{Q_2}\;\psi(\tan{Q_2})\left(P_2^2+P_3^2+\text{e}^{2Q_3}(P_1^2+c_1)+V_2(Q_2)+V_3(Q_3)\right),
\ee
where we have set $V_1(Q_1)=c_1$ to retain $P_1=e_1$ as a first integral.

\smallskip
This corresponds to a conformally flat metric in the $2-3$ space, defined by $P_1=\mbox{const}$, with $\text{e}^{2Q_3}(P_1^2+c_1)$ (times the conformal factor) corresponding to a potential term.  Since the conformal factor is a function of only $Q_2$, the momentum $P_3$ corresponds to a Killing vector (in 2D).

We have the following 6 conformal elements:
\bea
&&  {\cal T}_{e_1} = e^{Q_3} (P_2 \sin Q_2+P_3 \cos Q_2),\quad {\cal T}_{h_1} = 2 P_3,\quad {\cal T}_{f_1}=e^{-Q_3} (P_2 \sin Q_2-P_3 \cos Q_2),  \nn  \\[-1mm]
&&                             \label{e1h1f1-conf-e1}   \\[-1mm]
&&  {\cal T}_{e_2} = e^{Q_3} (P_2 \cos Q_2-P_3 \sin Q_2),\quad {\cal T}_{h_2} = -2 P_2,\quad {\cal T}_{f_2}= 2e^{-Q_3} (P_2 \cos Q_2+P_3 \sin Q_2), \nn
\eea
\end{subequations}
which satisfy the relations of $\mathfrak{g}_1 + \mathfrak{g}_2$ in Table \ref{Tab:g1234}, together with the algebraic constraints (\ref{e1h4-conf-constraints}).

The Hamiltonian (\ref{e1h1f1-Ht2e1}) has the involutive integrals $H, P_1$ and $G_1=P_3^2+\text{e}^{2Q_3}(P_1^2+c_1)+V_3(Q_3)$.  For the super-integrable cases, the additional integrals can be written in terms of the conformal elements (\ref{e1h1f1-conf-e1}).

\subsubsection{The Case $\psi(z)=\frac{z^2}{\alpha+\beta z^2}$}\label{2D-super-reduce-e1h1f1-e1-1}

With this $\psi$, the Hamiltonian (\ref{e1h1f1-Ht2e1}) takes the form of (\ref{2D-super-reduce-e1h1f1-e1-1H}).  Apart from $H$ and $P_1$, we have the quadratic integrals $G_1, G_2$, given below.
\begin{subequations}
\bea
H &=& \frac{\sin^2(2 Q_2)}{2(\alpha+\beta)+2(\alpha-\beta) \cos(2Q_2)} \, \left(P_2^2+P_3^2+e^{2 Q_3} (P_1^2+c_1)+c_2\csc^2{Q_2}+c_3 e^{4Q_3}\right), \label{2D-super-reduce-e1h1f1-e1-1H}  \\
G_1 &=& P_3^2+e^{2 Q_3}(P_1^2+c_1)+c_3 e^{4Q_3} ,   \label{2D-super-reduce-e1h1f1-e1-1G1}     \\
G_2 &=&  {\cal T}_{f_1}^2-\beta e^{-2 Q_3} \sec^2 Q_2\,H +c_3 e^{2Q_3}\cos^2{Q_2}.   \label{2D-super-reduce-e1h1f1-e1-1G2}
\eea
\end{subequations}
This Hamiltonian is the $D_4$ kinetic energy with a potential of ``type A" in the classification of \cite{03-11}, but with a slight variation of the type described in Remark \ref{rem:variations}.

\subsubsection*{In Cartesian Coordinates}

In the original Cartesian coordinates, the Hamiltonian is of the form
\be\label{2D-super-reduce-e1h1f1-e1-1qp}
H = \frac{q_2^2q_3^2}{\alpha q_2^2+\beta q_3^2}\left(p_1^2+p_2^2+p_3^2 +c_1+\frac{c_2}{q_3^2}+c_3(q_2^2+q_3^2)\right).
\ee
This is a conformally flat extension of case 2 in Table II of \cite{90-22}, constant curvature when either $\alpha=0$ or $\beta=0$.

\subsubsection{The Case $\psi(z)=\frac{\sqrt{1+z^2}}{\alpha \sqrt{1+z^2}  +\beta z}$}\label{2D-super-reduce-e1h1f1-e1-2}

With this $\psi$, the Hamiltonian (\ref{e1h1f1-Ht2e1}) takes the form of (\ref{2D-super-reduce-e1h1f1-e1-2H}).  Apart from $H$ and $P_1$, we have the quadratic integrals $G_1, G_2$, given below.  {\small
\begin{subequations}
\bea
H &=& \frac{\cos^2 Q_2}{\alpha+\beta \sin Q_2} \, \left(P_2^2+P_3^2+e^{2 Q_3}\left(P_1^2+c_1\right) +c_2\sec^2{Q_2}+c_3e^{Q_3}\right),  \label{2D-super-reduce-e1h1f1-e1-2H}\\
G_1 &=& P_3^2+e^{2 Q_3}(P_1^2+c_1)+c_3 e^{Q_3} ,  \label{2D-super-reduce-e1h1f1-e1-2G1}\\
G_2 &=& 2 P_2 {\cal T}_{f_1}-\frac{1}{2} e^{-Q_3}(3 \beta -\beta \cos (2 Q_2)+4 \alpha \sin Q_2)\, \sec^2 Q_2\,H +\sin{Q_2} (2c_2 e^{-Q_3}\sec^2{Q_2}+c_3).  \label{2D-super-reduce-e1h1f1-e1-2G2}
\eea
\end{subequations}   }
This Hamiltonian is the $D_4$ kinetic energy with a potential of ``type A" in the classification of \cite{03-11}.

\subsubsection*{Other Coordinate Systems}

In the original Cartesian coordinates, the Hamiltonian is of the form
\begin{subequations}
\be\label{2D-super-reduce-e1h1f1-e1-2qp}
H = \frac{q_2^2\sqrt{q_2^2+q_3^2}}{\alpha\sqrt{q_2^2+q_3^2}+\beta q_3}\left(p_1^2+p_2^2+p_3^2 +c_1+\frac{c_2}{q_2^2}+\frac{c_3}{\sqrt{q_2^2+q_3^2}}\right).
\ee
This is a conformally flat extension of case 4 in Table II of \cite{90-22}, constant curvature when $\beta=0$.

In {\em polar coordinates} $(q_1,q_2,q_3) = (r \cos\theta,r \sin\theta \cos\varphi ,r \sin\theta \sin\varphi)$, the general $H$ and $G_1$ take the form:
\bea
H &=& \frac{r^2 \sin^2\theta \cos^2\varphi}{\alpha+\beta \sin\varphi}\left(p_r^2+\frac{p_\theta^2}{r^2}+\frac{p_\varphi^2}{r^2\sin^2\theta}
                                                   +c_1 +\frac{c_2 \sec^2\varphi}{r^2 \sin^2\theta} +\frac{c_3}{r \sin\theta}\right),     \label{2D-super-reduce-e1h1f1-e1-2r}   \\
G_1 &=& (r^2 p_r^2+p_\theta^2) \sin^2\theta +c_1 r^2 \sin^2\theta +c_3 r\sin \theta.  \label{2D-super-reduce-e1h1f1-e1-2G1r}
\eea
In general, (\ref{2D-super-reduce-e1h1f1-e1-2r}) separates into only two components, in $(r,\theta)$ and $\varphi$.  The integral $G_1$ is related to this separability.  When $c_3=0$, the system fully separates in these coordinates.
\end{subequations}


\section{Systems with Isometry Algebra $\langle e_1,e_2,h_2\rangle$}\label{2D-super-reduce-e1e2h2}

In this (non-commutative) case we adapt coordinates to either $e_1\rightarrow P_1$ \underline{or} $h_2\rightarrow P_3$.  The case of $e_2$ is related to that of $e_1$ through the involution $\iota_{12}$.
The general Hamiltonian in this class is:
\be\label{3d-e1e2h2}
H = H_0 + U(q_1,q_2,q_3)  =  \psi(q_3) \left(p_1^2+p_2^2+p_3^2\right)  + U(q_1,q_2,q_3).
\ee

\subsection{Reduction with respect to $h_2\mapsto P_3$}

The generating function
\begin{subequations}
\be\label{e1e2h2-S3h2}
  S=\sqrt{q_1^2+q_2^2}\;P_1+q_3\;P_2-\frac12\arctan\left(\frac{q_1}{q_2}\right)\;P_3,
\ee
gives a Hamiltonian of type (\ref{H-12}):
\be\label{e1e2h2-Ht2h2}
  H= \psi(Q_2)\left(P_1^2+P_2^2+\frac1{4Q_1^2}(P_3^2+c_1)+V_1(Q_1)+V_2(Q_2)\right),
\ee
where we have set $V_3(Q_3)=c_1$ to retain $P_3=h_2$ as a first integral.

\smallskip
This case corresponds to a conformally flat metric in the $1-2$ space, defined by $P_3 = \mbox{const.}$, with $\frac1{4Q_1^2}(P_3^2+c_1)$ (times the conformal factor) corresponding to a potential term.  Since the conformal factor is a function of only $Q_2$, the momentum $P_1$ corresponds to a Killing vector (in 2D).

We have the following 6 conformal elements:
\bea
&&  {\cal T}_{e_1} = P_1,\;\;\; {\cal T}_{h_1} = -2(Q_1P_1+Q_2P_2),\;\;\; {\cal T}_{f_1}= (Q_2^2-Q_1^2) P_1-2 Q_1Q_2 P_2,  \nn  \\[-1mm]
&&    \label{e1e2h2-conf-h2}    \\[-1mm]
&&  {\cal T}_{e_2} = P_2,\quad {\cal T}_{h_2} = 2 (Q_1P_2-Q_2 P_1),\quad {\cal T}_{f_2}= 2(Q_1^2-Q_2^2) P_2-4 Q_1Q_2 P_1, \nn
\eea
which satisfy the relations of $\mathfrak{g}_1 + \mathfrak{g}_2$ in Table \ref{Tab:g1234}, together with the algebraic constraints (\ref{e1h4-conf-constraints}).
\end{subequations}

The Hamiltonian (\ref{e1e2h2-Ht2h2}) has the involutive integrals $H, P_3$ and $G_1=P_1^2+\frac1{4Q_1^2}(P_3^2+c_1)+V_1(Q_1)$.  For the super-integrable cases, the additional integrals can be written in terms of the conformal elements (\ref{e1e2h2-conf-h2}).

\subsubsection{The Case $\psi(z)=\frac{1}{\alpha z+\beta}$}\label{2D-super-reduce-e1e2h2-h2-1}

With this $\psi$, the Hamiltonian (\ref{e1e2h2-Ht2h2}) takes the form given below.  Apart from $H$ and $P_3$, we have the quadratic integrals $G_1, G_2$, given below.
\bea
&&  H = \frac{1}{\alpha Q_2+\beta} \, \left(P_1^2+P_2^2+ \frac{P_3^2+c_1}{4 Q_1^2}+c_2 (Q_1^2+4 Q_2^2)+c_3\right),\nn\\
&&  G_1=P_1^2+\frac{P_3^2+c_1}{4 Q_1^2}+c_2Q_1^2,\quad
  G_2=-2 P_2 {\cal T}_{h_1}-\left(4 (\beta+\alpha Q_2) Q_2+\alpha Q_1^2\right)\,H +8c_2(Q_1^2+2Q_2^2)+4c_3Q_2.  \nn
\eea
This is the $D_1$ kinetic energy with a potential of Case 1 in \cite{02-6}.

\subsubsection*{In Cartesian Coordinates}

In the original Cartesian coordinates, the Hamiltonian is of the form
\be\label{2D-super-reduce-e1e2h2-h2-1qp}
H = \frac1{\alpha q_3+\beta}\left(p_1^2+p_2^2+p_3^2 +\frac{c_1}{4(q_1^2+q_2^2)}+c_2(q_1^2+q_2^2+4q_3^2)+c_3\right).
\ee
When $\alpha=0$, this is just case 6 in Table II of \cite{90-22}.

\subsubsection{The Case $\psi(z)=\frac{z^2}{\alpha z^2+\beta}$}\label{2D-super-reduce-e1e2h2-h2-2}

With this $\psi$, the Hamiltonian (\ref{e1e2h2-Ht2h2}) has the conformal factor of Darboux-Koenigs metric $D_2$.  Apart from $H$ and $P_3$, we have the quadratic integrals $G_1, G_2$, given below.
\bea
&&  H = \frac{Q_2^2}{\alpha Q_2^2+\beta} \, \left(P_1^2+P_2^2+\frac{P_3^2+c_1}{4 Q_1^2}+c_2(Q_1^2+Q_2^2)+c_3\right),\nn\\
&&  G_1=P_1^2+\frac{P_3^2+c_1}{4 Q_1^2}+c_2Q_1^2,\qquad
  G_2=P_2 {\cal T}_{f_2}+2\left(\alpha Q_2^2-\frac{\beta Q_1^2}{Q_2^2}\right) H-2c_2(Q_1^2+Q_2^2)Q_2^2-2c_3Q_2^2.  \nn
\eea
This is the $D_2$ kinetic energy with a potential of type B in \cite{03-11}.

\subsubsection*{In Cartesian Coordinates}

In the original Cartesian coordinates, the Hamiltonian is of the form
\be\label{2D-super-reduce-e1e2h2-h2-2qp}
H = \frac{q_3^2}{\alpha q_3^2+\beta}\left(p_1^2+p_2^2+p_3^2 +\frac{c_1}{4(q_1^2+q_2^2)}+c_2(q_1^2+q_2^2+q_3^2)+c_3\right).
\ee
When $\beta=0$, this is just case 5 in Table II of \cite{90-22}.

\subsection{Reduction with respect to $e_1\mapsto P_1$}

Here we have the very simple case, of type (\ref{H-23}):
\begin{subequations}
\be \label{e1e2h2-Hte1}
Q_1=q_1,\;\;\; Q_2=q_3,\;\;\; Q_3=q_2 \quad\Rightarrow\quad  \tilde H = \psi(Q_2) \left(P_2^2+P_3^2+(P_1^2+c_1)+V_2(Q_2)+V_3(Q_3)\right),
\ee
where we have set $V_1(Q_1)=c_1$ to retain $P_1=e_1$ as a first integral.

\smallskip
This case corresponds to a conformally flat metric in the $2-3$ space, defined by $P_1 = \mbox{const.}$, with $(P_1^2+c_1)$ (times the conformal factor) corresponding to a potential term.  Since the conformal factor is a function of only $Q_2$, the momentum $P_3$ corresponds to a Killing vector (in 2D).

We have the following 6 conformal elements:
\bea
&&  {\cal T}_{e_1} = P_3,\;\;\; {\cal T}_{h_1} = -2(Q_2P_2+Q_3P_3),\;\;\; {\cal T}_{f_1}= (Q_2^2-Q_3^2) P_3-2 Q_2Q_3 P_2,  \nn  \\[-1mm]
&&    \label{e1e2h2-conf-e1}    \\[-1mm]
&&  {\cal T}_{e_2} = P_2,\quad {\cal T}_{h_2} = 2 (Q_3P_2-Q_2 P_3),\quad {\cal T}_{f_2}= 2(Q_3^2-Q_2^2) P_2-4 Q_2Q_3 P_3, \nn
\eea
which satisfy the relations of $\mathfrak{g}_1 + \mathfrak{g}_2$ in Table \ref{Tab:g1234}, together with the algebraic constraints (\ref{e1h4-conf-constraints}).
\end{subequations}

The Hamiltonian (\ref{e1e2h2-Hte1}) has the involutive integrals $H, P_1$ and $G_1=P_3^2+(P_1^2+c_1)+V_3(Q_3)$.  For the super-integrable cases, the additional integrals can be written in terms of the conformal elements (\ref{e1e2h2-conf-e1}).

\subsubsection{The Case $\psi(z)=\frac{1}{\alpha z+\beta}$}\label{2D-super-reduce-e1e2h2-e1-1}

With this $\psi$, the Hamiltonian (\ref{e1e2h2-Hte1}) takes the form given below.  Apart from $H$ and $P_1$, we have the quadratic integrals $G_1, G_2$, given below:
\bea
&&  H = \frac{1}{\alpha Q_2+\beta} \, \left(P_2^2+P_3^2+P_1^2+c_1+c_2\left(Q_2^2+Q_3^2\right)+2c_3Q_3\right),\nn\\
&&  G_1=P_3^2+c_2Q_3^2+2c_3Q_3,  \qquad G_2=P_2 P_3-\frac{1}{2}\,\alpha  Q_3\,\tilde{H}+c_2Q_2Q_3+c_3Q_2.  \nn
\eea
This is the $D_1$ kinetic energy with a potential of Case 2 in \cite{02-6}.

\subsubsection{The Case $\psi(z)=\frac{z^2}{\alpha z^2+\beta}$}\label{2D-super-reduce-e1e2h2-e1-2}

With this $\psi$, the Hamiltonian (\ref{e1e2h2-Hte1}) has the conformal factor of Darboux-Koenigs metric $D_2$.  Apart from $H$ and $P_1$, we have the quadratic integrals $G_1, G_2$, given below:
\bea
&&  H = \frac{Q_2^2}{\alpha Q_2^2+\beta} \, \left(P_2^2+P_3^2+ P_1^2+c_1+c_2 (Q_2^2+4 Q_3^2)+c_3Q_3\right),\nn\\
&&   G_1=P_3^2+4c_2Q_3^2+c_3Q_3,  \qquad G_2=-2P_2 {\cal T}_{h_2} +\frac{4 \beta Q_3}{Q_2^2}\,H+(4c_2Q_3+c_3)Q_2^2.  \nn
\eea
This is the $D_2$ kinetic energy with a potential of type A in \cite{03-11}.


\section{Systems with Isometry Algebra $\left<h_2,h_3,h_4\right>$}\label{2D-super-reduce-h2h3h4}

We give one transformation, corresponding to $h_4\rightarrow P_3$.  The transformations using either $h_2$ or $h_3$ are equivalent under the involutions $\iota_{13}$ and $\iota_{12}$, respectively.

The general Hamiltonian in this class is:
\be\label{3d-h2h3h4}
H = H_0 + U(q_1,q_2,q_3) = \psi\left(q_1^2+q_2^2+q_3^2\right)\, \left(p_1^2+p_2^2+p_3^2\right) + U(q_1,q_2,q_3).
\ee

\subsection{Reduction using the Isometry $h_4\mapsto P_3$}

The generating function
\begin{subequations}
\be\label{h2h3h4-S3}
  S= \arctan\left(\frac{\sqrt{q_2^2+q_3^2}}{q_1}\right)\;P_1+\frac12\log{(q_1^2+q_2^2+q_3^2)}\;P_2+\frac14\arctan\left(\frac{q_2}{q_3}\right)\;P_3,
\ee
gives a Hamiltonian of type (\ref{H-12}):
\be\label{h2h3h4-Ht2}
  H=\text{e}^{-2Q_2}\psi\left(\text{e}^{2Q_2}\right)\left(P_1^2+P_2^2+\frac{1}{16\sin^2{Q_1}}(P_3^2+c_1) +V_1(Q_1)+V_2(Q_2)\right),
\ee
where we have set $V_3(Q_3)=c_1$ to retain $P_3=h_4$ as a first integral.

\smallskip
This case corresponds to a conformally flat metric in the $1-2$ space, defined by $P_3 = \mbox{const.}$, with $\frac{1}{16\sin^2{Q_1}}(P_3^2+c_1)$ (times the conformal factor) corresponding to a potential term.  Since the conformal factor is a function of only $Q_2$, the momentum $P_1$ corresponds to a Killing vector (in 2D).

We have the following 6 conformal elements:
\bea
&&  {\cal T}_{e_1} = e^{Q_2} (P_1 \sin Q_1+P_2 \cos Q_1),\quad {\cal T}_{h_1} = 2 P_2,\quad {\cal T}_{f_1}= e^{-Q_2} (P_1 \sin Q_1-P_2 \cos Q_1),  \nn  \\[-1mm]
&&                             \label{h2h3h4-conf}   \\[-1mm]
&&  {\cal T}_{e_2} = e^{Q_2} (P_1 \cos Q_1-P_2 \sin Q_1),\quad {\cal T}_{h_2} = -2 P_1,\quad {\cal T}_{f_2}= 2e^{-Q_2} (P_1 \cos Q_1+P_2 \sin Q_1), \nn
\eea
which satisfy the relations of $\mathfrak{g}_1 + \mathfrak{g}_2$ in Table \ref{Tab:g1234}, together with the algebraic constraints (\ref{e1h4-conf-constraints}).
\end{subequations}

The Hamiltonian (\ref{h2h3h4-Ht2}) has the involutive integrals $H, P_3$ and $G_1=P_1^2+\frac{1}{16\sin^2{Q_1}}(P_3^2+c_1) +V_1(Q_1)$.  For the super-integrable cases, the additional integrals can be written in terms of the conformal elements (\ref{h2h3h4-conf}).

\subsubsection{The Case $\psi(z)=\frac{\sqrt{z}}{\alpha \sqrt{z}  +\beta}$}\label{h2h3h4-Ht2-h4-1}

With this $\psi$, the Hamiltonian (\ref{h2h3h4-Ht2}) takes the form given below.  Apart from $H$ and $P_3$, we have the quadratic integrals $G_1, G_2$, given below.
\begin{subequations}
\bea \label{Kepler-HGG-QP}
H &=& \frac{e^{-Q_2}}{\alpha e^{Q_2}+\beta} \, \left(P_1^2+P_2^2+\frac{P_3^2+c_1}{16 \sin^2 Q_1}+\frac{c_2\cos{Q_1}}{\sin^2{Q_1}}+c_3e^{Q_2}\right),\label{Kepler-H-QP}\\
 G_1  &=&  P_1^2+\frac{P_3^2+c_1}{16 \sin^2 Q_1}+\frac{c_2\cos{Q_1}}{\sin^2{Q_1}},    \label{Kepler-G1-QP}\\
    G_2 &=&  2 P_2 {\cal T}_{f_1}+\left(2 \alpha e^{Q_2}+\beta\right)\,\cos Q_1 \,H+c_2e^{-Q_2}-c_3\cos{Q_1}.  \label{Kepler-G2-QP}
\eea
\end{subequations}
This is the $D_3$ kinetic energy with the potential of ``type B" in \cite{03-11}.

\br[Spherically symmetric case]
When $c_1=c_2=0$, this Hamiltonian is invariant under the 3D isometry algebra
$$
\langle 2\sin{(4Q_3)}P_1+\frac12\cos{(4Q_3)}\cot{Q_1}P_3,2\cos{(4Q_3)}P_1-\frac12\sin{(4Q_3)}\cot{Q_1}P_3,P_3\rangle,
$$
which is just the original isometry algebra $\langle h_2,h_3,h_4\rangle$, expressed in these coordinates.  In this case, $G_1$ reduces to the Casimir function;
$$
G_1 =  P_1^2+\frac{P_3^2}{16 \sin^2 Q_1} = \frac{1}{4}\, \left(h_2^2+h_3^2+\frac{1}{4} h_4^2\right).
$$
\er

\subsubsection*{Relationship with the Kepler Problem}

The Hamiltonian (\ref{Kepler-H-QP}) is related to the Kepler problem, which is better seen in polar coordinates, which we approach via the original {\underline{Cartesian coordinates}, given by the inverse canonical transformation, generated by (\ref{h2h3h4-S3}):  {\small
\begin{subequations}
\bea \label{Kepler-HGG-qp}
    H &=& \frac{\sqrt{q_1^2+q_2^2+q_3^2}}{\alpha \sqrt{q_1^2+q_2^2+q_3^2}+\beta}\left(p_1^2+p_2^2+p_3^2 +\frac{1}{q_2^2+q_3^2} \left(\frac{c_1}{16}+\frac{c_2 q_1}{\sqrt{q_1^2+q_2^2+q_3^2}}\right) +\frac{c_3}{\sqrt{q_1^2+q_2^2+q_3^2}}\right),\label{Kepler-H-qp}\\
 G_1  &=&  L^2 + \frac{1}{q_2^2+q_3^2} \left(\frac{1}{16}\, c_1 (q_1^2+q_2^2+q_3^2)+c_2 q_1\, \sqrt{q_1^2+q_2^2+q_3^2}\right),    \label{Kepler-G1-qp}\\
    G_2 &=&  e_1 h_1+q_1 \left(\frac{\beta+2 \alpha\, \sqrt{q_1^2+q_2^2+q_3^2}}{\sqrt{q_1^2+q_2^2+q_3^2}}\right)\, H+\frac{c_2-q_1 c_3}{\sqrt{q_1^2+q_2^2+q_3^2}} ,  \label{Kepler-G2-qp}
\eea
\end{subequations}  }
{\flushleft where $L^2 = \frac{1}{4}\, \left(h_2^2+h_3^2+\frac{1}{4} h_4^2\right)$ is the usual Casimir of the rotation algebra, and $G_2$ is just an extension of the first component of Runge-Lenz vector.  When $c_1=c_2=\beta=0, \, \alpha=1$, we have exactly (twice) the first component of    }
$$
{\bf K} = {\bf p} \times {\bf L} + \frac{\frac{1}{2}c_3 {\bf q}}{\sqrt{q_1^2+q_2^2+q_3^2}}, \quad\mbox{where}\quad {\bf L} = \left\{-\frac{1}{4}\, h_4,-\frac{1}{2}\, h_3,\frac{1}{2}\, h_2\right\}.
$$
In this rotationally invariant case, the remaining components of the Runge-Lenz vector arise via $\{h_2,G_2\}$ and $\{h_3,G_2\}$.

Since we have reduced with respect to $h_4$, we must choose polar coordinates which make $h_4$ proportional to $p_\varphi$, so use the coordinate transformation;  $(q_1,q_2,q_3) = (r \cos \theta,r \sin\theta \cos \varphi,r \sin\theta \sin \varphi)$, leading to
\begin{subequations}\label{Kepler-HG-r}
\bea
 H &=& \frac{r}{\alpha r+\beta}\left(p_r^2+\frac{p^2_{\theta}}{r^2}+\frac{p^2_{\varphi}}{r^2\sin^2{\theta}}+\frac{c_1}{16r^2\sin^2{\theta}} +\frac{c_2\cos{\theta}}{r^2\sin^2{\theta}}+\frac{c_3}{r}\right),  \label{Kepler-H-r}\\
 G_1 &=& p_{\theta}^2+\frac{p^2_{\varphi}}{\sin^2{\theta}}+\frac{c_1+c_3\cos{\theta}}{\sin^2{\theta}},\label{Kepler-G1-r}\\
  G_2 &=& 2 p_r (p_\theta \sin \theta-r p_r \cos\theta)+(2\alpha r+\beta)\cos{\theta}\;H+\frac{c_2}{r} -c_3\cos{\theta}, \label{Kepler-G2-r}
\eea
\end{subequations}
which is clearly an extension of the Kepler problem and is spherically symmetric when $c_1=c_2=0$.

\br
As a consequence, we see that the combined transformation, which is just $(Q_1,Q_2,Q_3)= \left(\theta,\log r,\frac{1}{8}(\pi-2\varphi)\right)$ gives a direct relation between the Darboux-Koenigs Hamiltonian (\ref{Kepler-H-QP}) and the generalised Kepler problem (\ref{Kepler-H-r}).
\er

\subsubsection{The Case $\psi(z)=\frac{1}{\alpha z  +\beta}$}\label{h2h3h4-Ht2-h4-2}

With this $\psi$, the Hamiltonian (\ref{h2h3h4-Ht2}) takes the form given below.  Apart from $H$ and $P_3$, we have the quadratic integrals $G_1, G_2$, given below.
\begin{subequations}
\bea
  H &=& \frac{e^{-2Q_2}}{\alpha e^{2Q_2}+\beta} \, \left(P_1^2+P_2^2+\frac{P_3^2+c_1}{16 \sin^2 Q_1}+\frac{c_2}{\sin^2{(2Q_1)}}+c_3 e^{2Q_2}\right),\label{h2h3h4-h4-2-H-QP}\\
  G_1 &=& P_1^2+\frac{P_3^2+c_1}{16 \sin^2 Q_1}+\frac{c_2}{\sin^2{(2Q_1)}},   \label{h2h3h4-h4-2-G1-QP}   \\
  G_2 &=& {\cal T}_{f_1}^2 -\alpha e^{2Q_2}\,\cos^2 Q_1\,\, H+\frac{c_2e^{-2Q_2}}{4\cos^2{Q_1}}.  \label{h2h3h4-h4-2-G2-QP}
\eea
\end{subequations}
This is the $D_3$ kinetic energy with the potential of ``type B" in \cite{03-11}.

\subsubsection*{Other Coordinate Systems}

We again return to the original {\underline{Cartesian coordinates}, given by the inverse canonical transformation, generated by (\ref{h2h3h4-S3}):
\begin{subequations}
\bea
  H &=& \frac{1}{\alpha(q_1^2+q_2^2+q_3^2)+\beta}\left(p_1^2+p_2^2+p_3^2 +\frac{c_1+4 c_2}{16(q_2^2+q_3^2)}+\frac{c_2}{4q_1^2}+c_3\right),\label{h2h3h4-h4-2-H-qp}\\
  G_1 &=& L^2+\frac{(c_1+4 c_2)q_1^2}{16 (q_2^2+q_3^2)}+\frac{4 c_2 (q_2^2+q_3^2)}{q_1^2},   \label{h2h3h4-h4-2-G1-qp}   \\
  G_2 &=& p_1^2-\alpha q_1^2 H +\frac{c_2}{4 q_1^2},  \label{h2h3h4-h4-2-G2-qp}
\eea
where, again, $L^2 = \frac{1}{4}\, \left(h_2^2+h_3^2+\frac{1}{4} h_4^2\right)$ is the usual Casimir of the rotation algebra.
\end{subequations}

The transformation $(q_1,q_2,q_3) = (r \cos \theta,r \sin\theta \cos \varphi,r \sin\theta \sin \varphi)$ gives the system in polar coordinates:
\begin{subequations}\label{h2h3h4-h4-2-HG-r}
\bea
 H &=& \frac1{\alpha r^2+\beta}\left(p_r^2+\frac{p^2_{\theta}}{r^2}+\frac{p^2_{\varphi}}{r^2\sin^2{\theta}}+\frac{c_1}{16 r^2\sin^2{\theta}} +\frac{c_2}{r^2\sin^2{2\theta}}+c_3\right),   \label{h2h3h4-h4-2-H-r}\\
  G_1 &=& p_{\theta}^2+\frac{p^2_{\varphi}}{\sin^2{\theta}}+\frac{c_1}{16 \sin^2{\theta}} +\frac{c_2}{\sin^2{2\theta}},   \label{h2h3h4-h4-2-G1-r}\\
    G_2 &=&   \left(\cos{\theta}\;p_r-\frac{\sin{  \theta}}{r}\;p_{\theta}\right)^2-\alpha\;r^2\cos^2{\theta}\, H+\frac{c_2}{4 r^2\cos^2{\theta}}, \label{h2h3h4-h4-2-G2-r}
\eea
in which $H$ is separable.  It can be seen that $G_1$ is related to separability in these coordinates.
\end{subequations}

\section{Conclusions}\label{conclude}

In this paper we have extended the results of \cite{f20-2} to include potentials.  In fact, we initially worked within the {\em universal coordinates}, described in Section \ref{univ-coords}, in which the {\em kinetic energy} separates.  Therefore, we initially added a general separable potential, subject only to being invariant with respect to the ``reducing isometry''.  In these universal coordinates, in which the kinetic energies take Darboux-Koenigs form, we naturally reproduced the potentials found in \cite{02-6,03-11}.  However, since our 2D reductions are actually 3D Hamiltonians, we can change coordinates to find some remarkable connections with other well known super-integrable systems.  Clearly, the most interesting of these is the relation to the Kepler problem.  Not only do we obtain some general, non-flat versions of the Kepler Hamiltonian, but also generalised Runge-Lenz integrals, which are simply related to the integrals of some Darboux-Koenigs Hamiltonians.

An obvious question is whether, by breaking this connection with separability, we can obtain interesting potentials related to {\em higher order} first integrals.  Some third order integrals were considered in \cite{f19-3}, where one of the cases of \cite{11-3} was constructed, but even in 2 degrees of freedom this was complicated.  This general problem is discussed in \cite{17-4}, for systems in 2 degrees of freedom, with one Killing vector.

The quantum case in 2 degrees of freedom was presented in \cite{f19-3}, but the approach has not yet been adapted to 3 and higher degrees of freedom.  Super-integrability in the quantum case is closely related to degeneracy of eigenvalues \cite{14-2,f07-1,f18-3,f20-3}, with commuting operators being used to construct eigenfunctions with the same eigenvalue.

The general classical problem in higher degrees of freedom is quite open, although there certainly exist some interesting classes in the literature \cite{09-10}, including extensions of some of the examples presented in this paper.

\subsection*{Acknowledgements}

This work was supported by NSFC Grant No. 11871396 and NSF of Shaanxi Province of China, Grant No. 2018JM1005.


\end{document}